\documentclass[longauth]{aa} 

\usepackage{graphicx}
\usepackage{txfonts}
\usepackage{color}

\begin{document}

   \title{Gravitational slopes, geomorphology, and material strengths of the nucleus of comet 67P/Churyumov-Gerasimenko from OSIRIS observations}

   \author{
O.~Groussin\inst{1}
\and
L.~Jorda\inst{1}
\and
A.-T.~Auger\inst{1,2}
\and
E.~Kührt\inst{3}
\and
R.~Gaskell\inst{4}
\and
C.~Capanna\inst{1}
\and
F.~Scholten\inst{3}
\and
F.~Preusker\inst{3}
\and
P.~Lamy\inst{1}
\and
S.~Hviid\inst{3}
\and
J.~Knollenberg\inst{3}
\and
U.~Keller\inst{3,5}
\and
C.~Huettig\inst{3}
\and
H.~Sierks\inst{6}\and C.~Barbieri\inst{7} \and R.~Rodrigo\inst{8,9}\and D.~Koschny\inst{10} \and H.~Rickman\inst{11,12} 
\and
M.~F.~A'Hearn\inst{13} \and J.~Agarwal\inst{6} \and M.~A.~Barucci\inst{14} \and J.-L.~Bertaux\inst{15} \and I.~Bertini\inst{16} \and S.~Boudreault\inst{6} \and G.~Cremonese\inst{17} \and V.~Da Deppo\inst{18} \and B.~Davidsson\inst{11} \and S.~Debei\inst{17} \and M.~De Cecco\inst{19} \and M.~R.~El-Maarry\inst{20} \and S.~Fornasier\inst{14} \and M.~Fulle\inst{21} \and P.~J.~Guti\'errez\inst{22} \and C.~G\"uttler\inst{6} \and W.-H Ip\inst{23} \and J.-R.~Kramm\inst{6} \and M.~K\"uppers\inst{24} \and M.~Lazzarin\inst{7} \and L.~M.~Lara\inst{22} \and J.~J.~Lopez Moreno\inst{22} \and S.~Marchi\inst{25} \and F.~Marzari\inst{7} \and M. Massironi\inst{16,26} \and H.~Michalik\inst{27} \and G.~Naletto\inst{16,18,28} \and N.~Oklay\inst{6} \and A.~Pommerol\inst{20} \and M.~Pajola\inst{16} \and N.~Thomas\inst{20} \and I.~Toth\inst{29} \and C.~Tubiana\inst{6} \and J.-B.~Vincent\inst{6} 
          }

   \institute{Aix Marseille Universit\'e, CNRS, LAM (Laboratoire d'Astrophysique de Marseille) UMR 7326, 13388, Marseille, France 
        \and
        Laboratoire GEOPS (G\'eosciences Paris Sud), Bat. 509, Universit\'e Paris Sud, 91405 Orsay Cedex, France 
        \and
         Institute of Planetary Research, DLR, Rutherfordstrasse 2, 12489, Berlin, Germany
         \and
         Planetary Science Institute, Tucson (AZ), USA
         \and
         Institute for Geophysics and Extraterrestrial Physics, TU Braunschweig, 38106, Germany
         \and
         Max-Planck-Institut f\"ur Sonnensystemforschung, 37077 G\"ottingen, Germany
         \and
         Department of Physics and Astronomy, Padova University, Vicolo dell'Osservatorio 3, 35122, Padova, Italy
         \and
         Centro de Astrobiologia (INTA-CSIC), 28691 Villanueva de la Canada, Madrid, Spain 
         \and
         International Space Science Institute, Hallerstrasse 6, CH-3012 Bern, Switzerland 
         \and
         Scientific Support Office, European Space Agency, 2201, Noordwijk, The Netherlands
         \and
         Department of Physics and Astronomy, Uppsala University, Box 516, 75120, Uppsala, Sweden
         \and
         PAS Space Research Center, Bartycka 18A, PL-00716 Warszawa, Poland
         \and
         Department of Astronomy, University of Maryland, College Park, MD, 20742-2421, USA
         \and
         LESIA, Obs. de Paris, CNRS, Univ Paris 06, Univ. Paris-Diderot, 5 place J. Janssen, 92195 Meudon, France 
         \and
         LATMOS, CNRS/UVSQ/IPSL, 11 boulevard d'Alembert, 78280, Guyancourt, France
         \and
         Centro di Ateneo di Studi ed Attività Spaziali, "Giuseppe Colombo" (CISAS), University of Padova, via Venezia 15, 35131 Padova, Italy
         \and
         Department of Mech. Engineering University of Padova, via Venezia 1, 35131 Padova, Italy
         \and
         CNR-IFN UOS Padova LUXOR, via Trasea 7, 35131 Padova, Italy
         \and
         UNITN, Universit di Trento, via Mesiano, 77, 38100 Trento, Italy
         \and
         Physikalisches Institut, Sidlerstr. 5, University of Bern, CH-3012 Bern, Switzerland
         \and
         INAF - Osservatorio Astronomico, Via Tiepolo 11, 34143, Trieste, Italy
         \and
         Instituto de Astrofisica de Andaluc\'ia (CSIC), Glorieta de la Astronom\'ia s/n, 18008 Granada, Spain
         \and
         Institute for Space Science, Nat. Central Univ., 300 Chung Da Rd., 32054, Chung-Li, Taiwan
         \and
         Operations Department, European Space Astronomy Centre/ESA, P.O. Box 78, 28691 Villanueva de la Canada, Madrid, Spain
         \and
         Southwest Research Institute, 1050 Walnut St., Boulder, CO 80302, USA
         \and
         INAF, Osservatorio Astronomico di Padova, 35122 Padova, Italy.
         \and
         Institut f\"ur Datentechnik und Kommunikationsnetze der TU Braunschweig, Hans-Sommer-Str. 66, 38106 Braunschweig, Germany
         \and
         University of Padova, Department of Information Engineering, Via Gradenigo 6/B, 35131 Padova, Italy
         \and
         Konkoly Observatory, Budapest H-1525, P.O. Box 67, Hungary
   }

   \date{Received -- ; accepted --}

\titlerunning{Gravitational slopes, geomorphology and material strengths of the nucleus of comet 67P}
\authorrunning{Groussin, Jorda, Auger et al.}

 
  \abstract
   {} 
   {We study the link between gravitational slopes and the surface morphology on the nucleus of comet 67P/Churyumov-Gerasimenko and provide constraints on the mechanical properties of the cometary material (tensile, shear, and compressive strengths).}
   {We computed the gravitational slopes for five regions on the nucleus that are representative of the different morphologies observed on the surface (Imhotep, Ash, Seth, Hathor, and Agilkia), using two shape models computed from OSIRIS images by the stereo-photoclinometry (SPC) and stereo-photogrammetry (SPG) techniques. We estimated the tensile, shear, and compressive strengths using different surface morphologies (overhangs, collapsed structures, boulders, cliffs, and Philae's footprint) and mechanical considerations.}
   {The different regions show a similar general pattern in terms of the relation between gravitational slopes and terrain morphology: i) low-slope terrains (0\,--\,20$^{\circ}$) are covered by a fine material and contain a few large ($>$10\,m) and isolated boulders, ii) intermediate-slope terrains (20\,--\,45$^{\circ}$) are mainly fallen consolidated materials and debris fields, with numerous intermediate-size boulders from $<$1\,m to 10\,m for the majority of them, and iii) high-slope terrains (45\,--\,90$^{\circ}$) are cliffs that expose a consolidated material and do not show boulders or fine materials. The best range for the tensile strength of overhangs is 3\,--\,15\,Pa (upper limit of 150\,Pa), 4\,--\,30\,Pa for the shear strength of fine surface materials and boulders, and 30\,--\,150\,Pa for the compressive strength of overhangs (upper limit of 1500\,Pa). The strength-to-gravity ratio is similar for 67P and weak rocks on Earth. As a result of the low compressive strength, the interior of the nucleus may have been compressed sufficiently to initiate diagenesis, which could have contributed to the formation of layers. Our value for the tensile strength is comparable to that of dust aggregates formed by gravitational instability and tends to favor a formation of comets by the accrection of pebbles at low velocities.}
   {}

   \maketitle
%

\section{Introduction}

Rosetta has been orbiting comet 67P/Churyumov-Gerasimenko (67P) since August 2014. The OSIRIS cameras \citep{Keller2007} onboard this spacecraft have acquired hundreds of images of the surface with an unprecedented spatial resolution down to the decimeter scale \citep{Sierks2015}. The images reveal a complex nucleus surface made of smooth and hummocky terrains that are partially
or entirely covered by dust or expose a consolidated material, pits, cliffs, and fractures from the hundred meter scale to the decimeter scale \citep{Thomas2015}. The nature and origin of these terrains and morphological features are far from being understood, but remain of paramount importance to better constrain the formation and evolution scenario of the nucleus of 67P and comets in general. This paper focuses on the link between the nucleus gravitational slopes and surface morphology to provide constraints on the nature of the cometary material and its mechanical properties in particular (tensile, shear, and compressive strengths).

Gravitational slopes have only been measured on three cometary nuclei so far, 9P/Tempel\,1 \citep{Thomas2007}, 81P/Wild\,2 \citep{Jorda2015}, and 67P \citep{Jorda2015}. While the slopes of 9P are between 0$^{\circ}$ and 35$^{\circ}$, those of 81P and 67P cover a much wider range from 0$^{\circ}$ to $>$90$^{\circ}$, slopes exceeding 90$^{\circ}$ indicate overhangs. Beyond the different spatial resolution of the shape models used to compute the gravitational slopes for these three bodies, the differences between 9P on one side and 81P and 67P on the other side are interpreted as an aging effect by \citet{Jorda2015}. Following the scenario
described by these authors, comets that have spent more time in the inner solar system like 9P have been smoothed and have a narrower range of gravitational slopes than comets that have spent less time in the inner solar system like 81P and 67P. This planation process of the nucleus was also proposed by \citet{Basilevsky2007} for comets 19P/Borrelly, 81P, and 9P.

The tensile, shear, and compressive strengths of the cometary material have been estimated by several methods, including the Deep Impact experiment, comet breakup observations, laboratory experiments, and theoretical modeling (Table\,1). \citet{Biele2009} compiled and discussed these different estimates in an excellent review paper. From the Deep Impact experiment, the shear strength was estimated to be $<$65\,Pa by \citet{AHearn2005}, but might be any value between 0 and 12 kPa according to \citet{Holsapple2007};
this is an uncertainty of three orders of magnitude. From comet breakup, the tensile strength was estimated to be 5\,Pa by \citet{Asphaug1996} from the encounter of comet Shoemaker-Levy\,9 with Jupiter, to 100\,Pa for a sun-grazing comet with a radius of 1\,km \citep{Klinger1989}. \citet{Toth2006} and \citet{Davidsson2001} estimated that a tensile strength of $<$100\,Pa and 1-53\,Pa, respectively, is sufficient to keep a comet nucleus stable against its rotational breakup. From laboratory experiments, the compressive strength of cometary material analogs (water ice and dust mixture) was estimated to
be between 20\,kPa and 1\,MPa \citep{Jessberger1989,BarNun2007}, while the tensile strength was estimated to be 200\,--\,1100\,Pa for homogeneous SiO$_2$ dust samples \citep{Blum2006} and down to 1\,Pa for dust-aggregate samples \citep{Blum2014}. From modeling, the tensile strength of fluffy silicate dust/ice material was estimated to be 270\,Pa by \citet{Greenberg1995} and to be 5000\,Pa by \citet{Biele2009} for the same material made of water ice alone. Again from modeling, the compressive strength was estimated to be 6500 Pa for porous icy grains \citep{Sirono2000}.

Depending on whether the material is consolidated or unconsolidated, these strength estimates (Table\,1) depend, or not, on the scale at which they were measured. While the strength of unconsolidated material might be scale invariant, that of consolidated material follows a typical $d^{-q}$ power law, where $d$ is the scale and $q$ is the power exponent, with q$\sim$0.6 for water ice \citep{Petrovic2003}. From the kilometer to the millimeter scale, the strength of consolidated material can thus change by three orders of magnitude, still less than the above ranges, which cover up to five orders of magnitude. The different estimates clearly are {\it a priori} difficult to reconcile with each other, and large uncertainties remain on the tensile, shear, and compressive strengths of the cometary material.

Section 2 presents the data, shape models, and methods used in this paper. Section 3 discusses the link between gravitational slopes and surface morphologies for different types of terrains and regions on the nucleus. In Sect.\,4 we estimate the tensile, shear, and compressive strengths of the cometary material. Discussions and conclusions are presented in Sect.\,5.

\section{Data, shape models, and gravitational slopes}

All the images shown in this paper were acquired with the Narrow Angle Camera (NAC) of the Optical, Spectroscopic and Infrared Remote Imaging System (OSIRIS) onboard Rosetta \citep{Keller2007} since August 2014. Their spatial resolution varies between 18\,cm\,pix$^{-1}$ and 1.8\,m\,pix$^{-1}$.

For this work we used two different shape models of the nucleus of 67P, computed from OSIRIS images. The first shape model was computed by \citet{Jorda2015} using the stereo-photoclinometry technique (SPC). The second shape model was computed by \citet{Preusker2015} using the stereo-photogrammetry technique (SPG). The SPC shape model was resampled to match the resolution of the SPG shape model. The Digital Terrain Models (DTMs) extracted from the SPC and SPG shape models have a horizontal sampling of 2\,m and a typical vertical accuracy at the decimeter scale. 

We computed the local gravitational slopes for these two shape models, including the effects of the nucleus rotation and assuming a uniform density inside the nucleus. Details on the method are provided in \citet{Jorda2012}. The error on the gravitational slope is estimated to be 5$^{\circ}$. When not specified, the term \emph{\textup{slope}} in this paper always refers to the \emph{\textup{gravitational slope}}.

\section{Relation between gravitational slopes and surface morphologies}

\begin{figure*}
\centering
\includegraphics[width=\hsize]{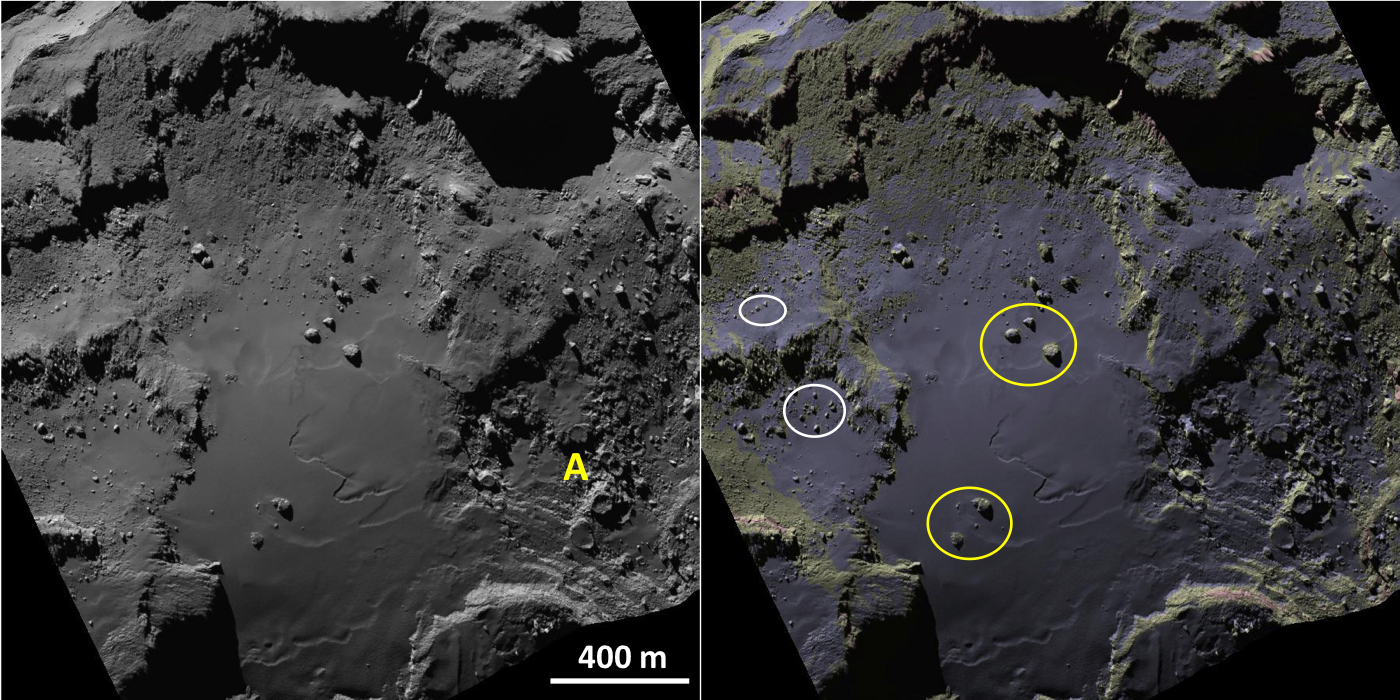}
\caption{{\it Left panel:} view of the Imhotep region. The letter A indicates the region of roundish features, observed near the gravitational low of the region. Image NAC\_2014-08-25T23.12.54 (spatial resolution: 0.94\,m\,pix$^{-1}$). {\it Right panel:} gravitational slopes of the Imhotep region, computed from the SPC shape model \citep{Jorda2015}, superimposed on the background image of the left panel. Blue corresponds to terrains with slope angles between 0$^{\circ}$ and 20$^{\circ}$, yellow to terrains with slope angles between 20$^{\circ}$ and 45$^{\circ}$ , and red to terrains with slope angles between 45$^{\circ}$ and 90$^{\circ}$. Yellow circles show examples of large and isolated boulders on slopes lower than 15$^{\circ}$. White circles show examples of boulder fields that end on slopes lower than 15$^{\circ}$.}
\label{Fig_Imhotep}
\end{figure*}

\begin{figure*}
\includegraphics[width=\hsize]{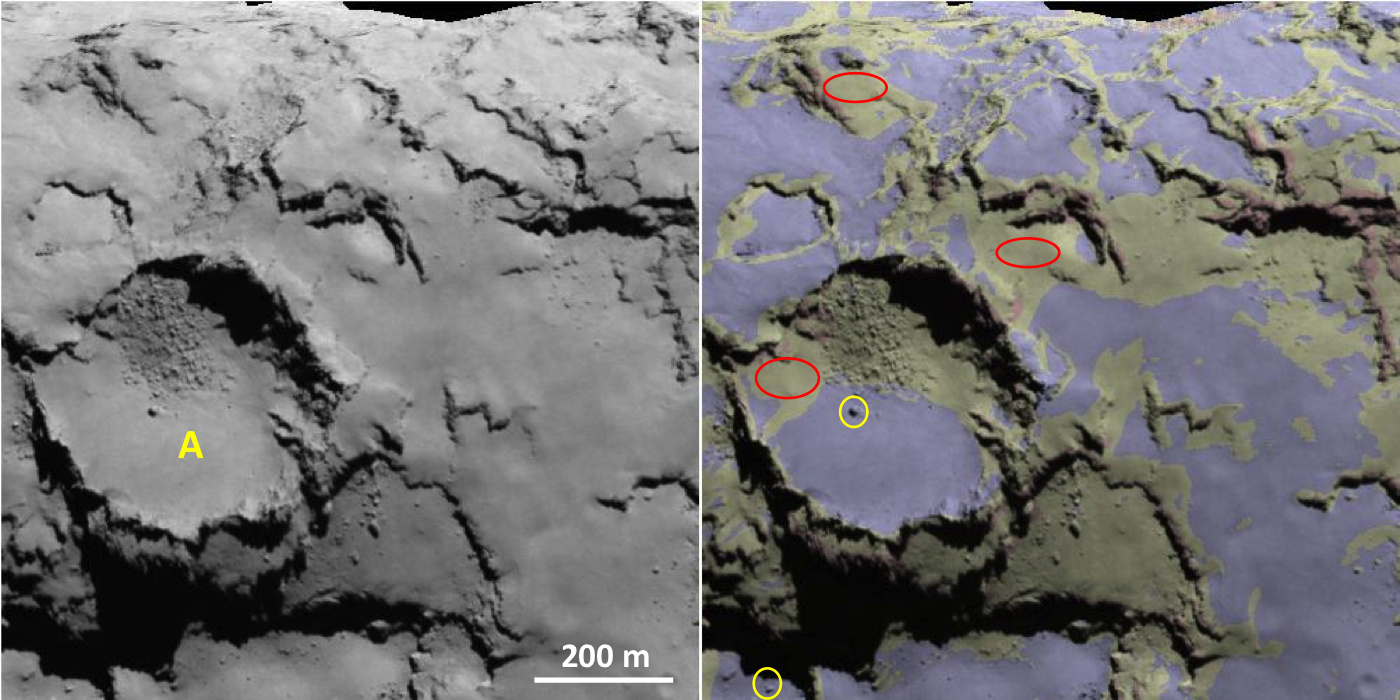}
\caption{{\it Left panel:} view of the Ash region. The letter A indicates the largest depression of the region. Image NAC\_2014-08-07T18.20.34 (spatial resolution: 1.5\,m\,pix$^{-1}$). {\it Right panel:} gravitational slopes of the Ash region, computed from the SPC shape model \citep{Jorda2015}, superimposed on the background image of the left panel. Blue corresponds to terrains with slope angles between 0$^{\circ}$ and 20$^{\circ}$, yellow to terrains with slope angles between 20$^{\circ}$ and 45$^{\circ}$ , and red to terrains with slope angles between 45$^{\circ}$ and 90$^{\circ}$. Yellow circles show examples of large and isolated boulders on slopes lower than 15$^{\circ}$. Red circles show examples of smooth terrains on slopes in the range 25\,--\,30$^{\circ}$.}
\label{Fig_Ash}
\end{figure*}

\begin{figure*}
\includegraphics[width=\hsize]{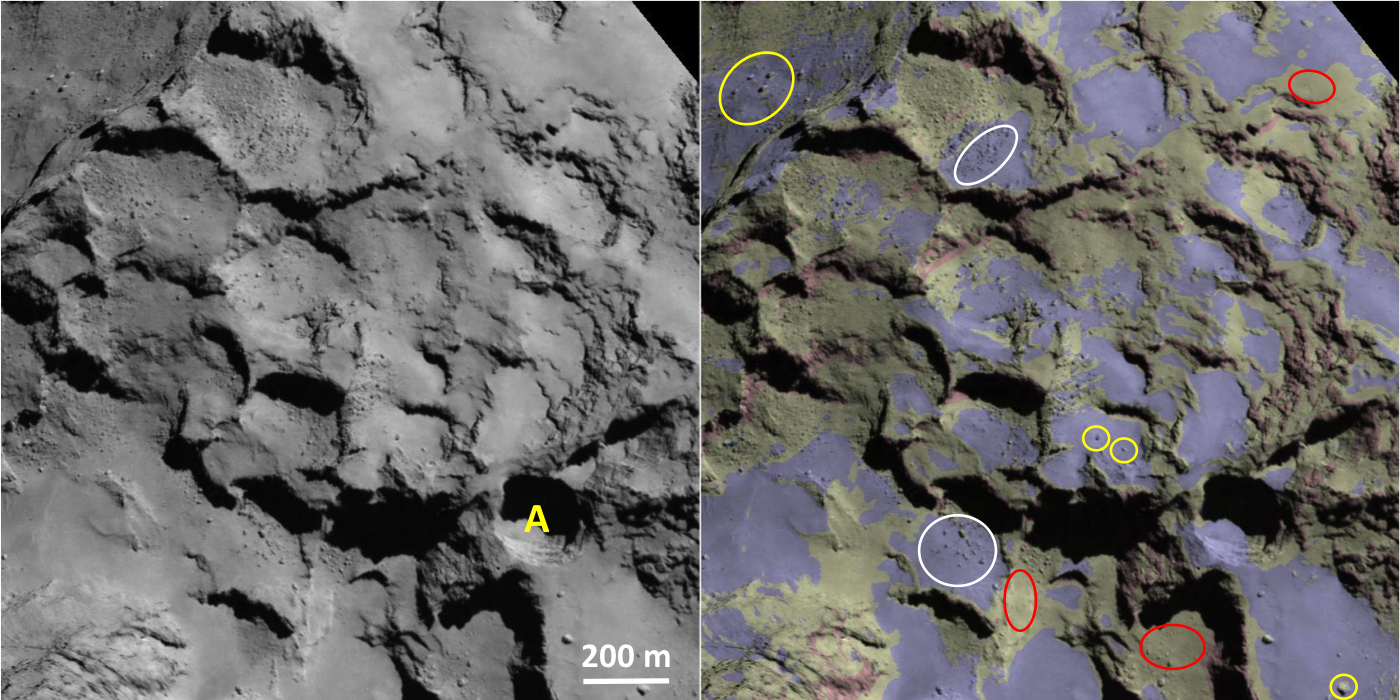}
\caption{{\it Left panel:} view of the Seth region. The letter A indicates the largest pit of the region. Image NAC\_2014-08-16T10.59.16 (spatial resolution: 1.8\,m\,pix$^{-1}$). {\it Right panel:} gravitational slopes of the Seth region, computed from the SPC shape model \citep{Jorda2015}, superimposed on the background image of the left panel. Blue corresponds to terrains with slope angles between 0$^{\circ}$ and 20$^{\circ}$, yellow to terrains with slope angles between 20$^{\circ}$ and 45$^{\circ}$ , and red to terrains with slope angles between 45$^{\circ}$ and 90$^{\circ}$. Yellow circles show examples of large and isolated boulders on slopes lower than 15$^{\circ}$. Red circles show examples of smooth terrains on slopes in the range 25\,--\,30$^{\circ}$. White circles show examples of boulder fields that end on slopes lower than 15$^{\circ}$.}
\label{Fig_Seth}
\end{figure*}

\begin{figure*}
\centering
\includegraphics[width=\hsize]{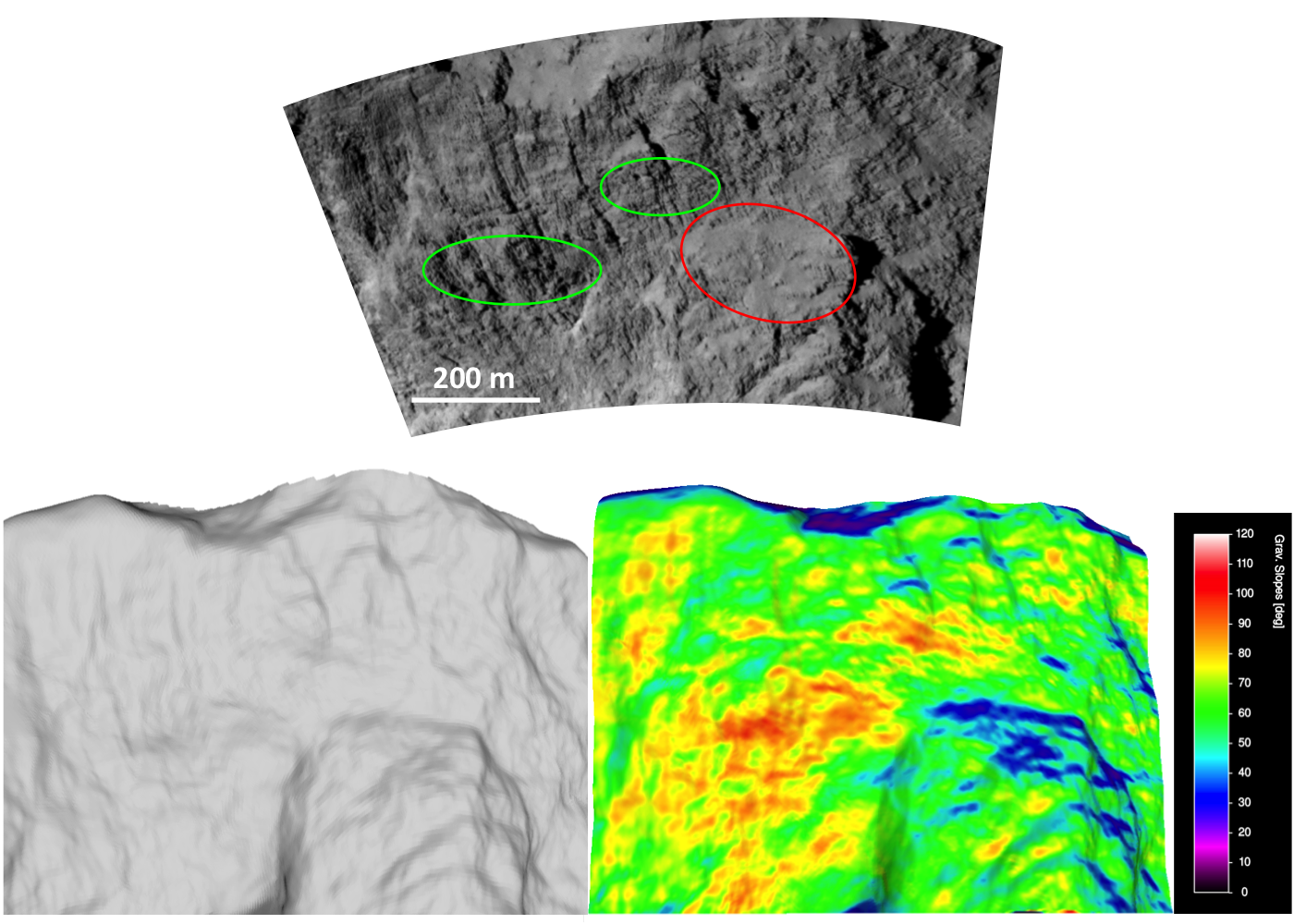}
\caption{{\it Upper panel:} View of the Hathor region. Image NAC\_2014-08-07T20.20.34 (spatial resolution: 1.5\,m\,pix$^{-1}$). The red circle shows an example of a smooth terrain on slopes in the range 25\,--\,30$^{\circ}$. Green circles show examples of overhangs with slopes $>$90$^{\circ}$. {\it Lower panel:} Extracted Digital Terrain Model ({\it left}) of the Hathor region with the corresponding gravitational slopes ({\it right}) computed from the SPC shape model \citep{Jorda2015}. For technical reasons related to the DTM extraction, we were not able to produce a figure for Hathor like Figs.\,\ref{Fig_Imhotep}, \ref{Fig_Ash}, \ref{Fig_Seth}, or \ref{Fig_Agilkia}.}
\label{Fig_Hathor}
\end{figure*}

\begin{figure*}
\includegraphics[width=\hsize]{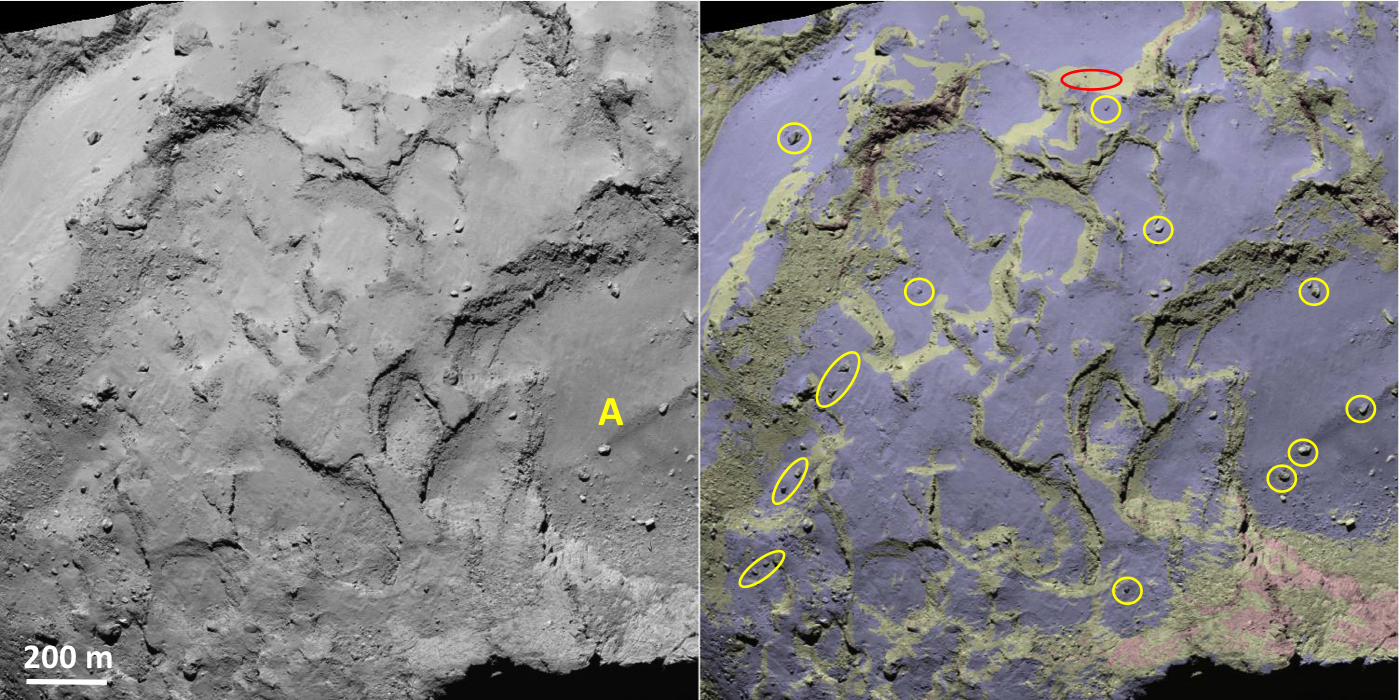}
\caption{{\it Left panel:} view of the Agilkia region. The letter A indicates the largest depression of the region. Image NAC\_2014-08-25T15.42.54 (spatial resolution: 0.94\,m\,pix$^{-1}$). {\it Right panel:} gravitational slopes of the Agilkia region, computed from the SPC shape model \citep{Jorda2015}, superimposed on the background image of the left panel. Blue corresponds to terrains with slope angles between 0$^{\circ}$ and 20$^{\circ}$, yellow to terrains with slope angles between 20$^{\circ}$ and 45$^{\circ}$ , and red to terrains with slope angles between 45$^{\circ}$ and 90$^{\circ}$. Yellow circles show examples of large and isolated boulders on slopes lower than 15$^{\circ}$. The red circle shows an example of a smooth terrain on slopes in the range 25\,--\,30$^{\circ}$.}
\label{Fig_Agilkia}
\end{figure*}

The geomorphology of the nucleus surface is diverse and constrained by several processes related to gravity and cometary activity \citep{Sierks2015,Thomas2015}. The link between the different types of terrains (smooth, hummocky, consolidated material, dust covered, with or without boulders, etc.) and their gravitational slope is important to better constrain the processes in play and the nature of the cometary material. In this section we study this relationship for five regions on the nucleus that are representative of the different morphologies observed on the surface. We refer to \citet{ElMaarry2015a} for the definition of the regions.
\begin{enumerate}
\item The Imhotep region (Fig.\,\ref{Fig_Imhotep}) -- This region presents a wide variety of terrains and morphologies. The most remarkable ones are the smooth terrains, the largest of which extend over 0.8\,km$^2$ , and the roundish features observed near the gravitational low of the region.\vspace*{0.2cm}
\item The Ash region (Fig.\,\ref{Fig_Ash}) -- This region is mostly covered by dust that is spatially unresolved at the decimeter scale. It shows several debris fields that are made of boulders, which are located at the feet of steep walls exposing a consolidated material. A large depression of 370\,m width is visible on the left side (noted A in Fig.\,\ref{Fig_Ash}).\vspace*{0.2cm}
\item The Seth region (Fig.\,\ref{Fig_Seth}) -- This region is dominated by circular depressions, most of them being accumulation basins with an opening toward a lower basin. Debris accumulates at the feet of the steep walls of basins. A large ($\sim$200\,m) and deep pit dominates the bottom right part of the region (noted A in Fig.\,\ref{Fig_Seth}).\vspace*{0.2cm} 
\item The Hathor region (Fig.\,\ref{Fig_Hathor}) -- This region is dominated by cliffs with a maximum height of $\sim$900\,m, exposing a consolidated material. Many anisotropies (e.g., fractures) are visible, organized in two main directions roughly perpendicular \citep{Thomas2015}. 
\vspace*{0.2cm}
\item The Agilkia region (Fig.\,\ref{Fig_Agilkia}) -- Agilkia is a large ``super region'' corresponding to the nominal landing site and includes the regions of Hatmehit, Ma'at, Nut, and Maftet. The Agilkia region is dominated by a very large depression of 800\,m width on its right side (noted A in Fig.\,\ref{Fig_Agilkia}). This region shows a variety of exposed consolidated material, dust-covered material, and boulder fields.
\end{enumerate}

\begin{figure*}
\centering
\includegraphics[width=9cm]{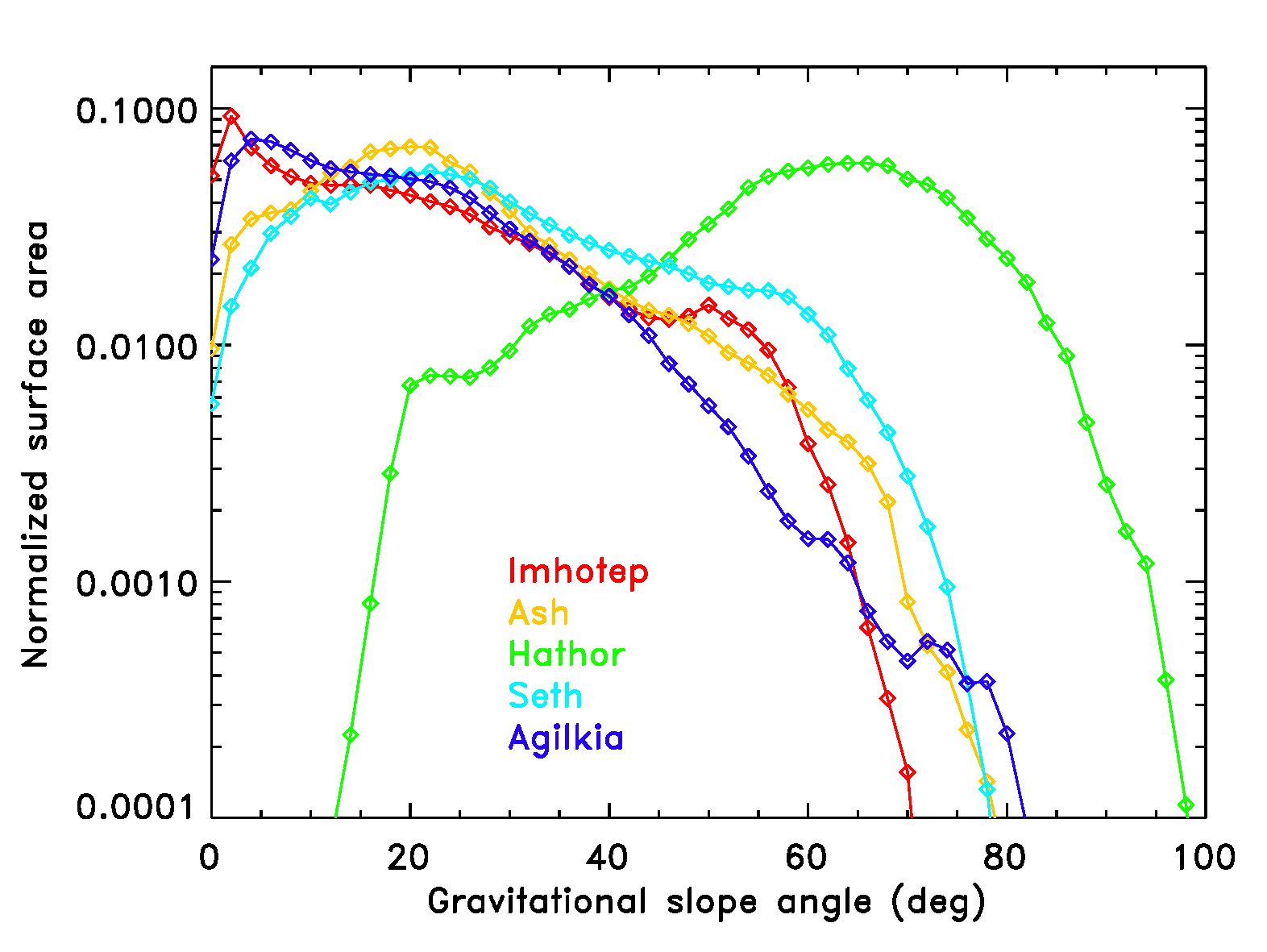}
\includegraphics[width=9cm]{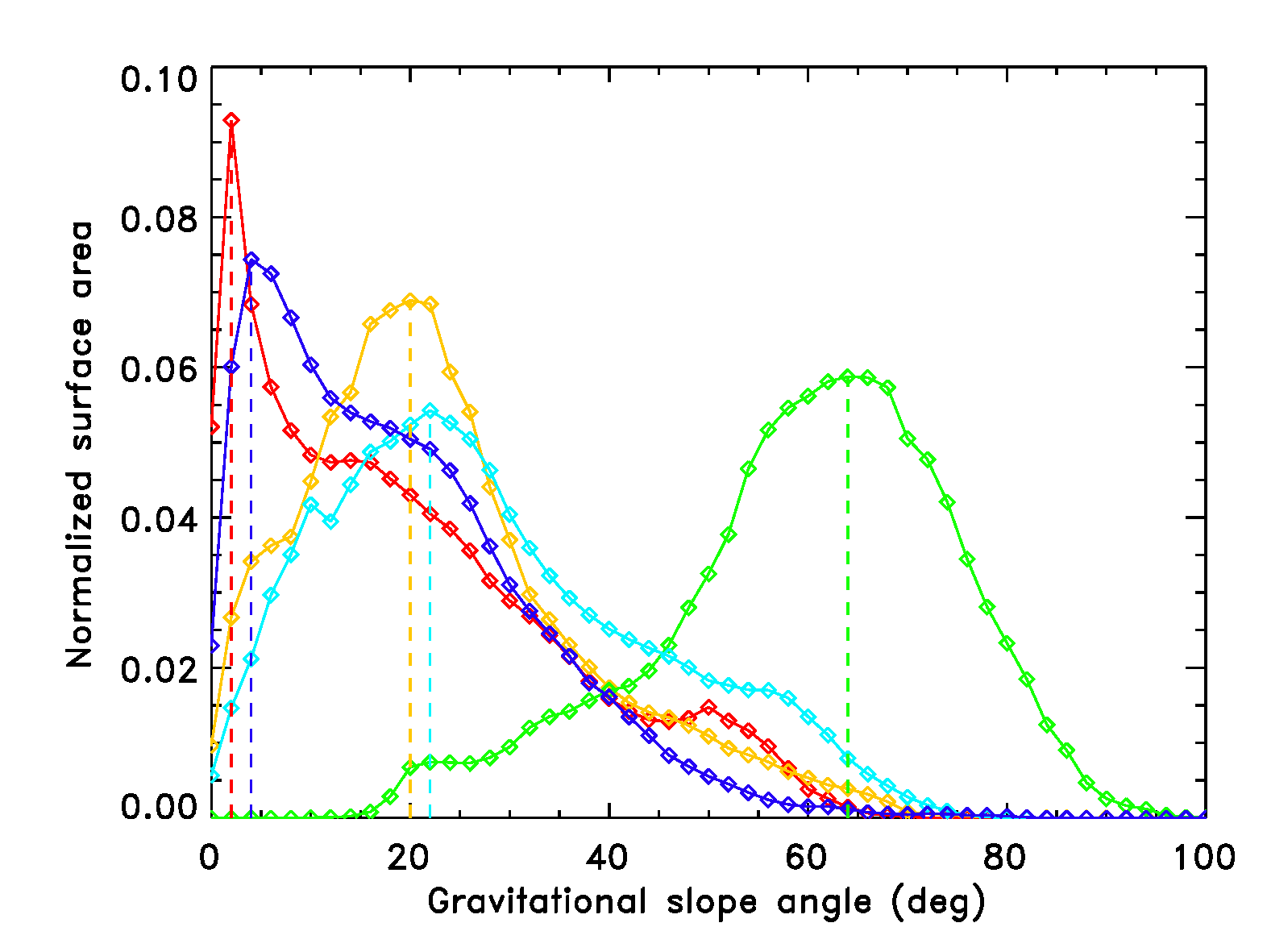}
\caption{Histogram of gravitational slope angles for the five regions, with one color per region, in logarithmic scale ({\it left}) and linear scale ({\it right}) for the Y axis. The values are normalized so that the sum of all Y values equals 1. Each region has a unique distribution of gravitational slopes, related to its unique topography. All regions cover a widee range of slopes from 0$^{\circ}$ to almost 100$^{\circ}$. The gravitational slopes were derived from the SPC shape model \citep{Jorda2015}.}
\label{Fig_histogram_slope}
\end{figure*}

The histogram of gravitational slope angles computed for each region is shown in Fig.\,\ref{Fig_histogram_slope}. Each region has a unique distribution of gravitational slopes, related to its unique topography. There are similarities between the different distributions, however. All regions cover a wide range of slopes from 0$^{\circ}$ to almost 100$^{\circ}$. All distributions show a peak (Fig.\,\ref{Fig_histogram_slope}, right panel). The peak of the distribution varies between 2$^{\circ}$ for Imhotep and 64$^{\circ}$ for Hathor. The peak value is a good indication of the overall flatness of the terrain, lower values indicating a flatter terrain on average. Secondary peaks, bumps, or shoulders are also visible; they are related to steep walls (e.g., bump around 55$^{\circ}$ on Seth and Imhotep) or smooth terrains (e.g., shoulder around 20$^{\circ}$ on Imhotep and Agilkia). Slopes in excess of 90$^{\circ}$ are indicative of overhangs and are
only visible on the Hathor cliffs; overhangs cover 0.1\% of the total Hathor area.

\begin{figure*}
\centering
\includegraphics[width=\hsize]{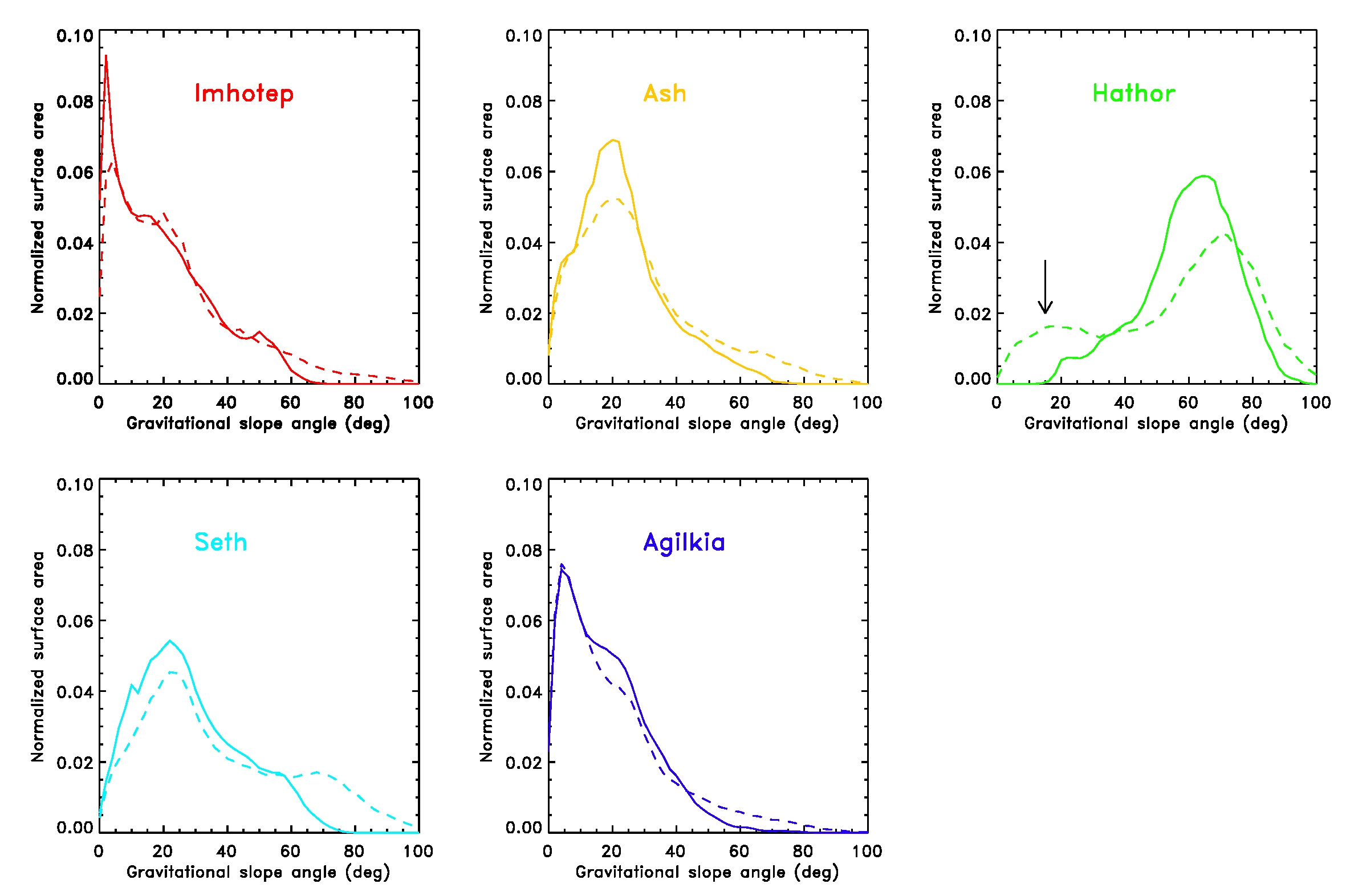}
\caption{Histograms of gravitational slope angles for the five regions, derived from the SPC shape model (solid line) and from the SPG shape model (dashed line). The black arrow on Hathor indicates a bump on the SPG slopes distribution that must be ignored since it results from the fact that the SPG DTM includes a large part of the Hapi region that is absent from the SPC DTM. The Hapi region, which is smooth and flat, thus introduces a bias in the distribution for low-slope angles.}
\label{Fig_histogram_slope_spc_spg}
\end{figure*}

Figure\,\ref{Fig_histogram_slope_spc_spg} allows a direct comparison between the SPC and SPG gravitational slopes. This is important to better estimate the error on the slope. The overall shape of the SPC and SPG slopes distributions is similar. In particular, the peaks are at the same position. The notable exception is Hathor, with a peak at 64$^{\circ}$ for the SPC slopes and 70$^{\circ}$ for the SPG slopes. This leads to a more general comment, which is that the slope distributions of SPC and SPG differ for slopes steeper than 60$^{\circ}$. The SPC shape model tends to systematically underestimate the fraction of the surface with steep slopes compared to the SPG shape model. This exercise shows that slopes are robust up to 60$^{\circ}$, with an error of $\pm$5$^{\circ}$ as given in Sect.\,2, but larger uncertainties affect slopes above 60$^{\circ}$, with an error up to 20$^{\circ}$ in some cases.

\begin{figure*}
\includegraphics[width=\hsize]{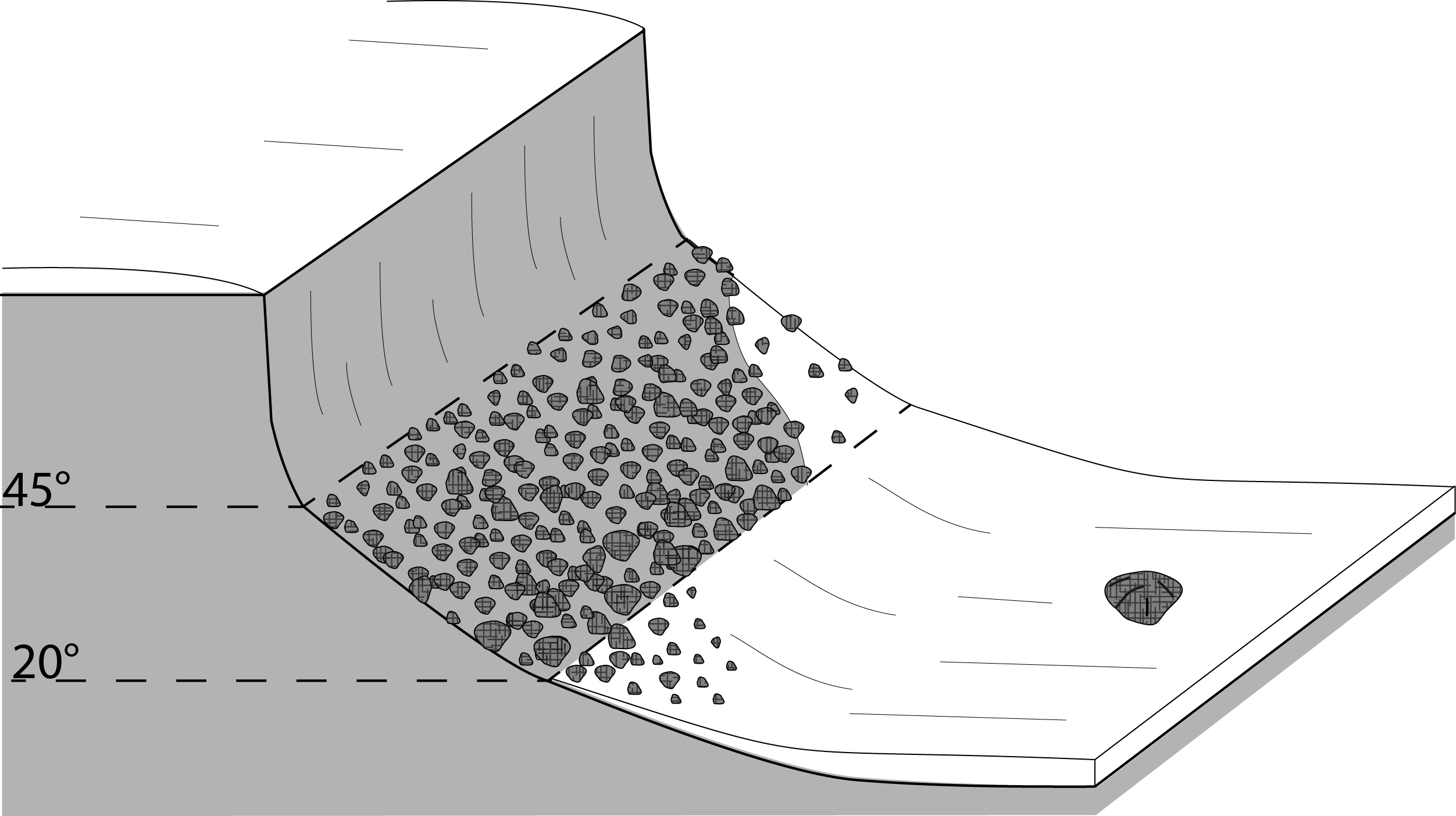}
\caption{Sketch of the general pattern in terms of relation between gravitational slopes and terrain morphologies: i) low-slope terrains (0\,--\,20$^{\circ}$) are covered by fine material and contain a few large ($>$10\,m) and isolated boulders, ii) intermediate-slope terrains (20\,--\,45$^{\circ}$) are mainly fallen consolidated materials and debris fields, with numerous intermediate-size boulders from $<$1\,m to 10\,m for the majority of them, and iii) high-slope terrains (45\,--\,90$^{\circ}$) are cliffs that
expose a consolidated material and do not show boulders or fine material. The borders between low- and intermediate-slope terrains are not always sharp (see text for details).}
\label{Fig_sketch_slope_terrain}
\end{figure*}

Figures\,\ref{Fig_Imhotep} - \ref{Fig_Agilkia} illustrate that all regions show a similar general pattern in terms of the relation between gravitational slope and terrain morphology. A sketch of this general pattern is shown in Fig.\,\ref{Fig_sketch_slope_terrain}:
\begin{itemize}
\item {\it Low-slope terrains.} Terrains with slopes in the range 0\,--\,20$^{\circ}$ are covered by a spatially unresolved material, that is, a material made of particles smaller than 20\,cm that we call the \emph{\textup{fine material}} in this paper. A few large isolated boulders are visible on these terrains, with a typical size larger than 10\,m. \vspace*{0.2cm}
\item {\it Intermediate-slope terrains.} Terrains with slopes in the range 20\,--\,45$^{\circ}$ are mainly fallen consolidated materials and debris fields, with numerous intermediate-size boulders from $<$1\,m to 10\,m for the majority of them \citep[see also][]{Pajola2015}. These terrains are covered by a dust deposit of variable thickness, which partially hides some boulders. Most of these terrains are located at the feet of high-slope terrains. \vspace*{0.2cm}
\item {\it High-slope terrains.} Terrains with slopes in the range 45\,--\,90$^{\circ}$ are cliffs that expose a consolidated material and do not show boulders or a fine material. These terrains probably show the bare nucleus.
\end{itemize}

This general pattern is very similar to what we find on Earth, particularly in young mountains like the Alps, where boulder fields are frequently observed at the feet of cliffs; theses
fields are
a result of cliff collapse. The sublimation of ices triggers erosion, most likely exacerbated by fractures; gravity controls the collapse like it does on Earth.

The presence of large boulders on slopes lower than 20$^{\circ}$ is intriguing (yellow circles in Figs.\,1 - \ref{Fig_Agilkia}). These boulders are usually isolated, far from high-slope terrains, and are large, tens of meters. They are too large to be lifted by gas drag resulting from regular cometary activity, this mechanism only applies to boulders smaller than the meter scale \citep{Groussin2003,AHearn2011,Kelley2013,Gundlach2015}. They could instead be air falls from outburst events, which are
limited in time, but much stronger in terms of released energy. Such events are sporadic and spatially localized, however; they may not be frequent enough to explain the presence of large boulders in all the regions we studied. An alternative solution is that large boulders are leftovers from previous basins and depression edges, when they were smaller and less eroded. 

The transition between intermediate- and high-slope terrains is sharp. This indicates that high-slope terrains have a steeper
slope than the angle of repose, which can be estimated to be $\theta_{\rm repose}=45\pm5^{\circ}$, a typical value for gravel on Earth \citep{Julien1995}. The transition between low- and intermediate-slope terrains is softer, with several examples of boulder fields on slopes lower than 20$^{\circ}$ (white circles in Fig.\,\ref{Fig_Seth}) and of smooth terrains on slopes steeper than 20$^{\circ}$ (red circles in Figs.\,\ref{Fig_Imhotep} - \ref{Fig_Agilkia}).

Boulder fields that end on slopes lower than 20$^{\circ}$ on Seth (white circles in Fig.\,\ref{Fig_Seth}) could result from a progressive eroding and degradation process of the nucleus. These boulder fields were probably rock falls, at the feet of the previous emplacement of cliffs, as the above isolated boulders. Following a slow degradation process that involves fractal fragmentation into small pieces and progressive covering by dust deposits, they now appear on low-slope terrains and are partially hidden \citep{Pajola2015}. The same degradation process can explain that a fine material of variable thickness partially hides boulder fields on intermediate-slope
terrains. 

Smooth terrains on slopes steeper than 20$^{\circ}$ (red circles in Figs.\,\ref{Fig_Imhotep} - \ref{Fig_Agilkia}) could be ancient boulder fields that were able to retain the dust deposit on a slope steeper than that of typical smooth terrains. The origin of the dust could be deposits from regular cometary activity or products of the degradation process of cliffs and boulders.

\section{Tensile, shear, and compressive strengths of the cometary material}

\subsection{Definition of strengths}

There are three types of strength for a given material: the tensile strength $\sigma_T$, the shear strength $\sigma_S$ , and the compressive strength $\sigma_C$.  They define the ability of a material to withstand mechanical constraints. Usually, $\sigma_T < \sigma_S < \sigma_C$. 
Depending on the nature of the material, consolidated or unconsolidated, the strength depends, or not, on the scale at which it is measured (Sect.\,1).
A review on scaling effects on structural strength can be found in \citet{Bazant1999}.

As explained in the introduction, large uncertainties remain in the values of $\sigma_T$, $\sigma_S$ , and $\sigma_C$ for the cometary material, sometimes of several orders of magnitude \citep[e.g.,][]{Biele2009}. We here intend to provide additional constraints on the strengths of the cometary material; more precisely, the tensile strength of overhangs and collapsed structures, the shear strength of fine materials, boulders and Hathor cliffs, and the compressive strength of fine materials and consolidated materials.

\subsection{Tensile strength of overhangs and collapsed structures}

\begin{figure*}
\includegraphics[width=10cm]{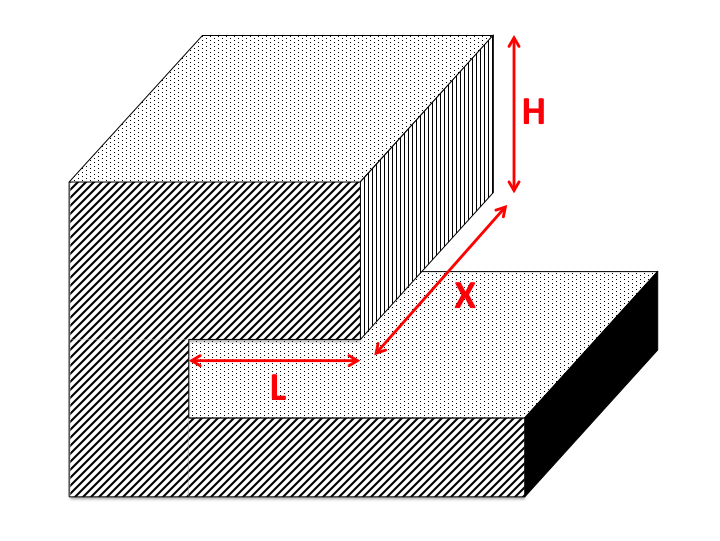}
\caption{Scheme of an overhang with rectangular shape $L \times X \times H$.}
\label{Fig_scheme_overhangs}
\end{figure*}

The tensile strength can be estimated from overhangs. From simple mechanics \citep[e.g.,][]{Tokashiki2010}, the failure of an overhang of rectangular shape due to bending will occur if the following condition is fulfilled:
\begin{equation}
\sigma_T < 3 \gamma \frac{L^2}{H}
\label{Eq_1}
\end{equation}
where $\gamma$ (N\,m$^{-3}$) is the unit weight, $L$ (m) the length of the overhang, and $H$ (m) its height as defined in Fig.\,\ref{Fig_scheme_overhangs}. Equation\,(\ref{Eq_1}), which is independent of the width $X$ of the overhang, can be translated into
\begin{equation}
\sigma_T < 3 \rho g \frac{L^2}{H}
\label{Eq_2}
\end{equation}
where $\rho=470$\,kg\,m$^{-3}$ is the density of the material (Sierks et al. 2015) and $g$ is the local gravity.

\begin{figure*}
\centering
\includegraphics[width=\hsize]{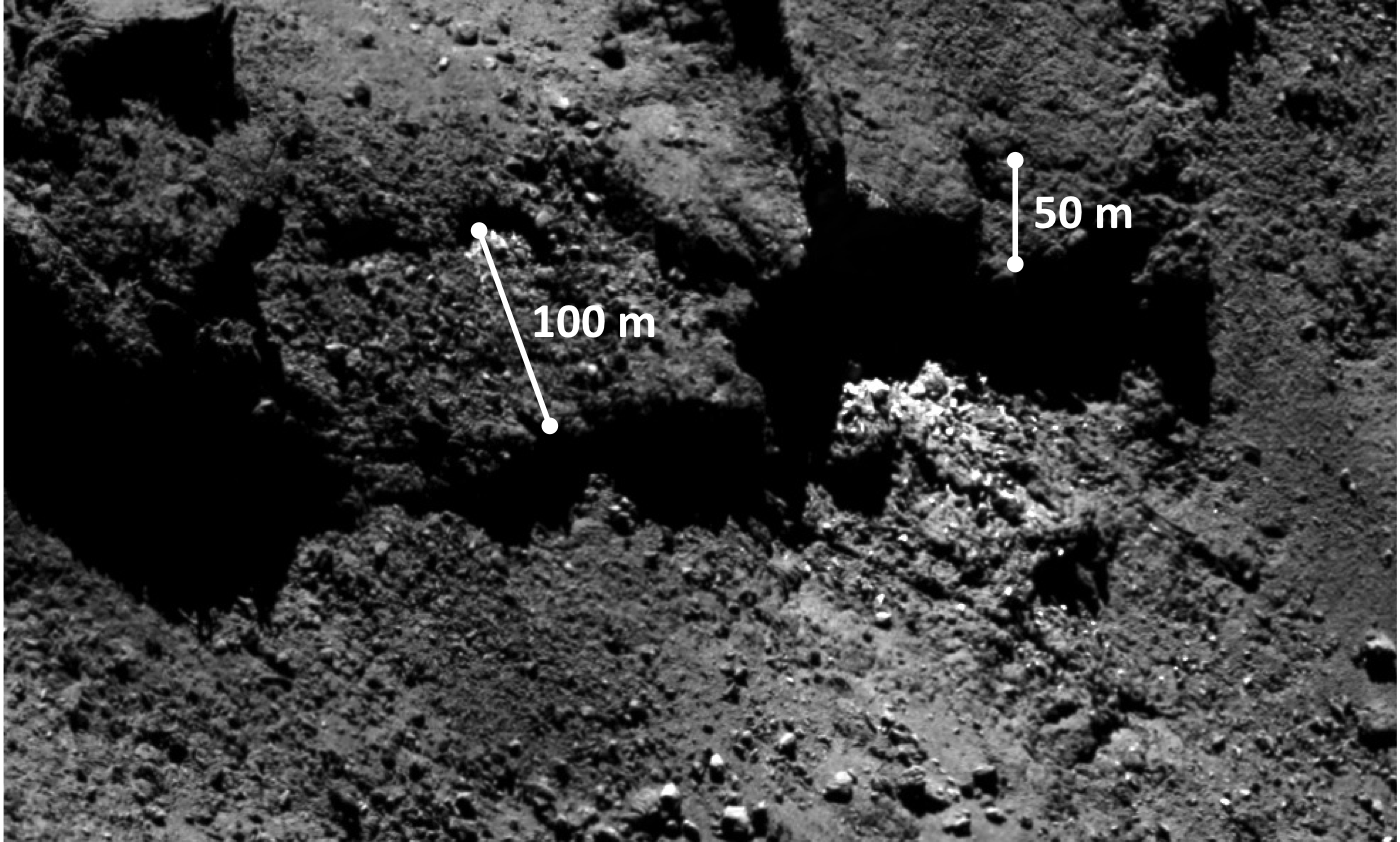}
\caption{Examples of large structures that already collapsed, with a length $L$ of 100\,m or 50\,m and a height $H$ of 30\,m (determined from the Digital Terrain Model). These structures are located at the northern border of the Imhotep region. Image NAC\_2014-09-16T15.44.07 (spatial resolution: 0.52\,m\,pix$^{-1}$).}
\label{Fig_image_overhang}
\end{figure*}

\begin{figure*}
\centering
\includegraphics[width=\hsize]{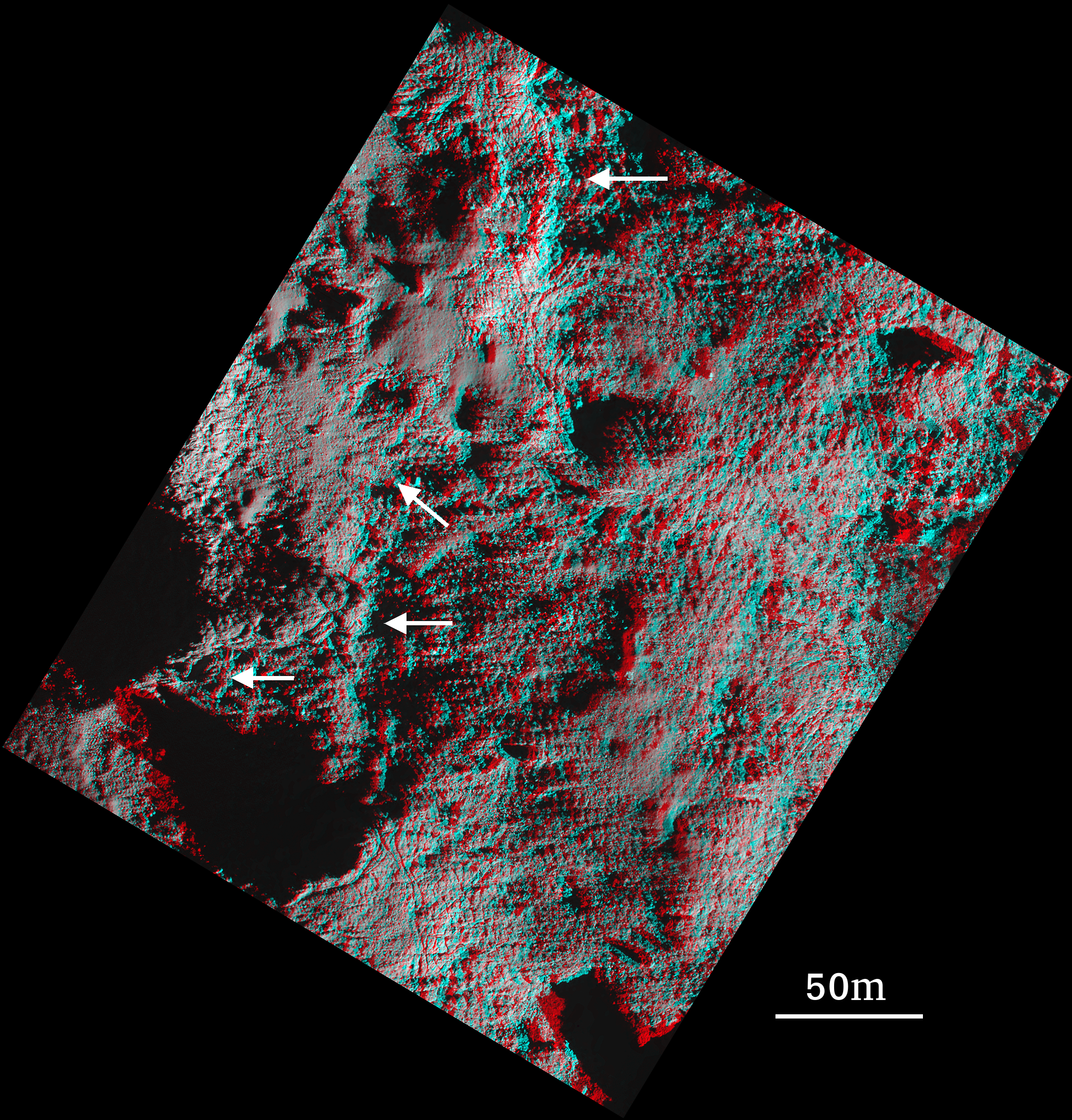}
\caption{Red/blue anaglyph of overhangs at a high spatial resolution of 18\,cm\,pix$^{-1}$, in the Maftet region. Several overhangs are visible in this anaglyph; they are indicated by white arrows, all with a length of $\sim$10\,m and a height of $\sim$5\,m. Image NAC\_2014-10-19T13.18.55 (spatial resolution: 0.18\,m\,pix$^{-1}$).}
\label{Fig_Anaglyph}
\end{figure*}

Several examples of overhangs or collapsed structures are shown in Figs.\,\ref{Fig_image_overhang} and\,\ref{Fig_Anaglyph}. Figure\,\ref{Fig_image_overhang} shows two large structures that collapsed, indicating that the tensile strength was exceeded. Using the DTM of the area, we estimated the length and height of the largest structure to be $L=100$\,m and $H=30$\,m. From Eq.\,(\ref{Eq_2}) with $g=2\times10^{-4}$\,m\,s$^{-2}$ , this gives an upper limit for the tensile strength $\sigma_T < 94$\,Pa. In Fig.\,\ref{Fig_Anaglyph}, several overhangs are visible with an estimated length of 10\,m and a height of 5\,m. The presence of mass wasting in the form of boulders at the feet of these overhangs indicates that they are close to breaking and thus good estimates of the tensile strength. This argument is also supported by the fact that some of these boulders have a size ($\sim$10\,m) similar to that of the overhangs themselves. From Eq.\,(\ref{Eq_2}) we derive $\sigma_T=5.6$\,Pa for these overhangs. 

This determination of the tensile strength is not accurate, mainly because of the uncertainty in the geometry (length and height). Taking these uncertainties into account, the tensile strength of small overhangs ($\sim$10\,m) is most likely in the range 3\,--\,15\,Pa, and the tensile strength of large collapsed structures ($\sim$100\,m) is lower than 150\,Pa.

\subsection{Shear strength of fine materials, boulders, and Hathor cliffs}

\begin{figure*}
\includegraphics[width=10cm]{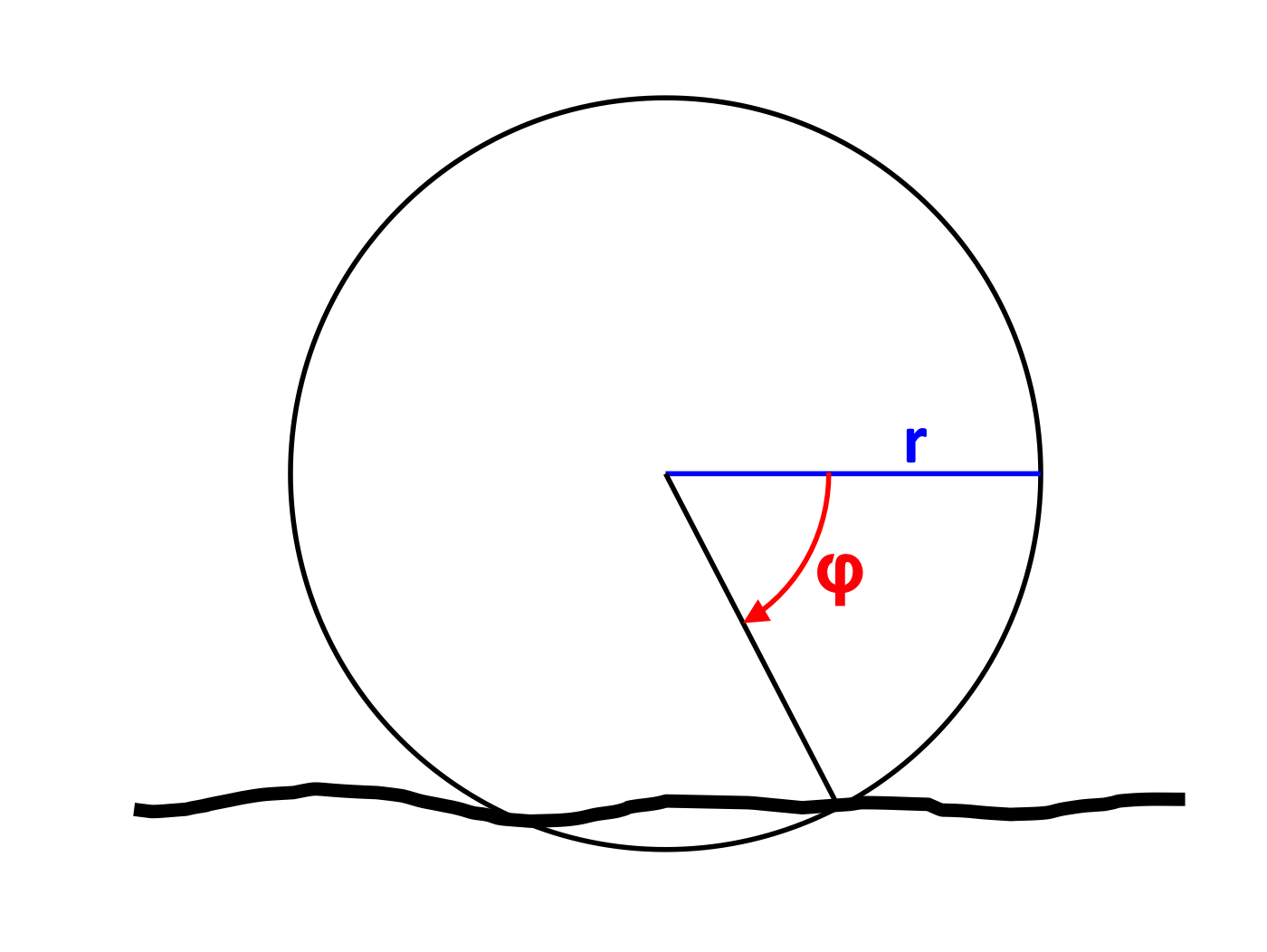}
\caption{Scheme of a boulder lying on the surface. The radius of the boulder is $r,$ and its contact area with the surface is defined by the angle $\varphi$.}
\label{Fig_scheme_boulder}
\end{figure*}

The shear strength can be calculated with Eq.\,(\ref{Eq_4}), where $\theta$ is the slope angle on which the boulder is located, $m$ (kg) is the mass of the considered boulder, and $A$ (m$^2$) is the contact area of the boulder with the terrain
underneath. The shear strength results from friction \emph{\textup{and}} cohesion, and we cannot separate these two physical quantities in this study.
\begin{equation}
\sigma_S = \frac{m g \sin \theta}{A} 
\label{Eq_4}
\end{equation}
For a boulder of radius $r$, Eq.\,(\ref{Eq_4}) translates into Eq.\,(\ref{Eq_5}) with $A=\pi (r \cos \varphi)^2$ (Fig.\,\ref{Fig_scheme_boulder}).
\begin{equation}
\sigma_S = \frac{\frac{4}{3}\pi r^3 \rho g \sin \theta}{\pi (r \cos \varphi)^2} = \frac{4 r \rho g \sin \theta}{3 \cos^2 \varphi}
\label{Eq_5}
\end{equation}

\begin{figure*}
\includegraphics[width=9cm]{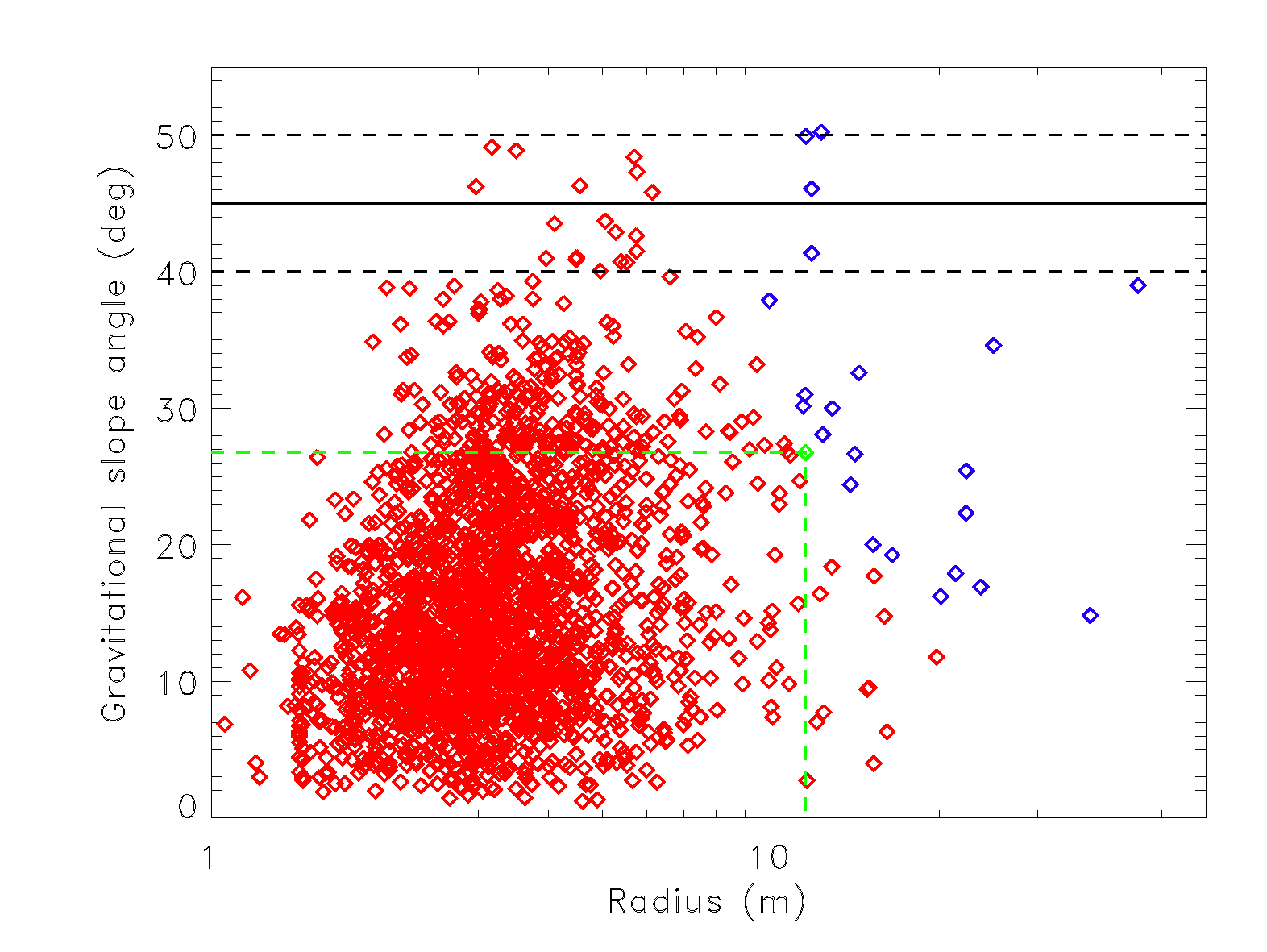}
\includegraphics[width=9cm]{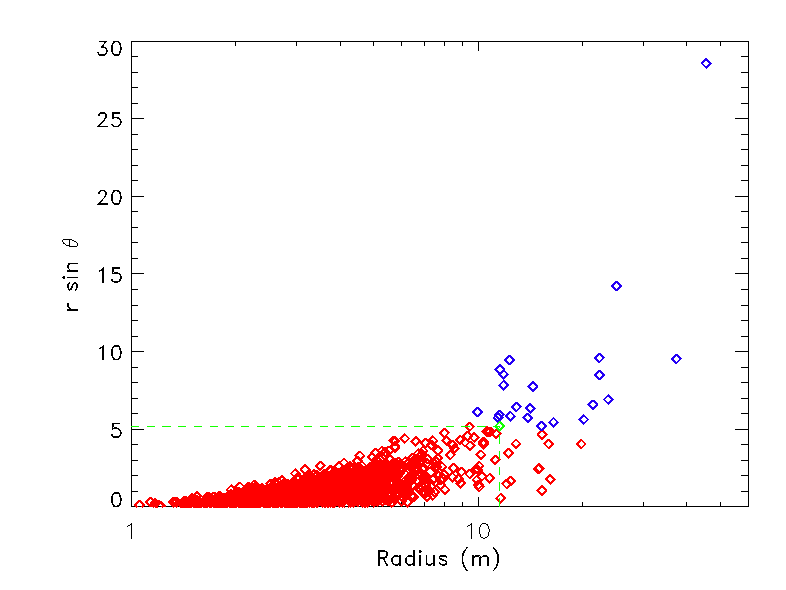}
\caption{{\it Left panel:} size (radius $r$) of 2207 boulders of the Imhotep region as a function of their gravitational slope angle $\theta$ \citep[adapted from][]{Auger2015}. The black horizontal lines correspond to the angle of repose $\theta_{\rm repose}=45\pm5^{\circ}$; there are no boulder on slopes larger than the angle of repose. {\it Right panel:} value $r \sin \theta$ for each boulder, as a function of its radius. {\it Both panels:} for robustness, we rejected the upper 1\% of the $r \sin \theta$ distribution (blue points). For the remaining 99\%, the boulder with the highest value $r \sin \theta$ is highlighted in green and corresponds to $r=11.5$\,m and $\theta=26.8^{\circ}$, as indicated by the green horizontal and vertical dashed lines.}
\label{Fig_stat_boulder}
\end{figure*}

The shear strength is constrained by the largest boulders on the highest gravitational slopes, more precisely. those with the highest $r \sin \theta$ value. From Fig.\,\ref{Fig_stat_boulder}, adapted from \citet{Auger2015}, the highest value for $r \sin \theta$ is 5.2\,m and corresponds to a boulder with radius $r=11.5$\,m located on a gravitational slope $\theta=26.8^{\circ}$. Assuming an area of contact of 1\,\% of the total boulder surface, defined with $\varphi=80^{\circ}$ in Eq.\,(\ref{Eq_5}), we obtain a shear strength $\sigma_S=22$\,Pa. Because of the uncertainty on the radius (2\,m), slope angle (5$^{\circ}$), and the fact that the area of contact may be larger than 1\,\% ($\varphi<$\,80$^{\circ}$), the shear strength of the boulders and fine material on which they stand is most likely in the range 4\,--\,30\,Pa. Figure\,\ref{Fig_stat_boulder} also illustrates that there are indeed no boulder on slopes steeper than the angle of repose $\theta_{\rm repose}=45\pm5^{\circ}$ (Sect.\,3).


A lower limit of the shear strength of the Hathor cliffs is provided by the lateral pressure at the bottom of these cliffs, given by Eq.\,(\ref{Eq_6}), where $h$ is the height of the cliff.
\begin{equation}
\sigma_S \geq \rho g h (1 - \sin \theta_{\rm repose})
\label{Eq_6}
\end{equation}
For a maximum height $h=900$\,m of the Hathor cliffs, this gives $\sigma_S \geq$\,30\,Pa. This value is an approximation since the cliff does not have a slope of 90$^{\circ}$ from top to bottom, meaning that it is not perfectly vertical, and the gravity changes from top to bottom. Nevertheless, it shows that due to the low gravity, a low strength is sufficient to withstand a high cliff.

\subsection{Compressive strength of fine and consolidated materials}

\begin{figure*}
\centering
\includegraphics[width=\hsize]{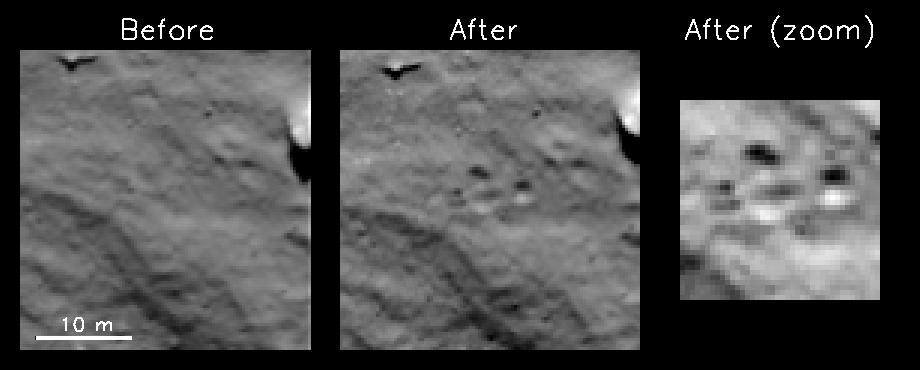}
\caption{Images of the first nominal landing site, before and after the Philae touchdown. The footprint of the three landing feet are clearly visible. Image NAC\_2014-11-12T15.18.52 and NAC\_2014-11-12T15.43.51 (spatial resolution: 0.31\,m\,pix$^{-1}$).}
\label{Fig_landing_site}
\end{figure*}

The compressive strength can be estimated from the footprints left by the lander Philae after it bounced on the nucleus surface at the first nominal landing site (Fig.\,\ref{Fig_landing_site}). We can derive the compressive strength with Eq.\,(\ref{Eq_7}), where $F$ (N) is the forced applied to a surface of area $A$ (m$^2$), $m$ (kg) is the mass of Philae, $v$ (m\,s$^{-1}$) is the lander impact velocity, and $d$ (m) is the depth at which the lander penetrated the surface material.
\begin{equation}
\sigma_C = \frac{F}{A} = \frac{\frac{1}{2}m v^2}{d A} 
\label{Eq_7}
\end{equation}

The depth of the largest footprint, which has a diameter of $\sim$5\,pix or 150\,cm, was estimated to be 20\,cm using a shape from shading analysis. This footprint was created by Philae when the first leg touched the ground; we assume here that the three legs did not touch the ground simultaneously. With a mass of 100\,kg for the lander, an impact velocity $v=1$\,m\,s$^{-1}$ and a contact area of 0.016\,m$^2$ per foot (each foot is made of two disks of 10\,cm diameter), we derive a compressive strength of 15.6\,kPa for the surface material. This value is an upper limit since the presence of footprints indicates that the impact process exceeded the compressive strength of the surface material. Moreover, as we discuss in Sect.\,5.4, the penetration of Philae of the surface materials was blocked by a hard layer under the surface, which also leads to overestimating the compressive strength.
 
A more reliable estimate can be given by taking into account mechanical considerations. In practice, the compressive strength of a consolidated material is always larger than its tensile strength by typically one order of magnitude (e.g., rocks like sandstone), so that we can reasonably assume $\sigma_C \sim 10 \sigma_T$ \citep{Sheorey1997}. This only applies to consolidated materials, however, such as overhangs or collapsed structures, but not to a unconsolidated material like the fine material on the surface \citep{Blum2004,Blum2006}. From the tensile strength calculated in Sect.\,4.2, we therefore derive a compressive strength of the consolidated material that most likely is in the range 30\,--\,150\,Pa, with an upper limit of 1.5\,kPa; this is well below the upper limit of 15.6\,kPa derived for the fine surface materials.

\section{Discussions and conclusions}

\subsection{Summary and comparison with other measurements}

Values for the tensile, shear, and compressive strengths are summarized in Table\,\ref{Table_strength}. Overall, the strengths are low and the cometary material can be considered as weak. Once scaled to the meter scale for comparison, our results agree
well with most estimates from other authors who used observations, laboratory experiments, or modeling, in particular for the tensile strength. This scaling should be taken with caution,
however, since consolidated and unconsolidated materials may not follow the same scaling law. Our estimate for the shear strength is at the lower end of the possible range mentioned by \citet{Holsapple2007}. Concerning the compressive strength, the discrepancy with \citet{Jessberger1989} results from the fact that they measured the compressive strength of a mixture made of liquid water and dust with a high fraction of water ($>$80\%). This mixture is cooled from room temperature to low temperature (253\,K, 233\,K, or 123\,K, depending on their test case), and water ice forms as compact hexagonal ice, known to be hard \citep[$\sigma_C=5$\,--\,25\,MPa;][]{Petrovic2003}. 

\subsection{Strength-to-gravity ratio}

The strengths are low on 67P, but the gravity is low as well. The ratio between the tensile strength at the 1\,m scale (7\,--\,34\,Pa) and gravity (2\,$\times$\,10$^{-4}$\,m\,s$^{-2}$) is 35\,--\,170\,$\times$\,10$^3$\,Pa\,s$^2$\,m$^{-1}$ on 67P. On Earth, weak rocks like siltstone, with a density of $\sim$2600\,kg\,m$^{-3}$ and a porosity of 21\,--\,41\,\%, have a typical tensile strength of 0.5\,MPa. For this type of weak rocks, the ratio between the tensile strength and Earth's gravity is 51\,$\times$\,10$^3$\,Pa\,s$^2$\,m$^{-1}$. This remarkable agreement between the strength-to-gravity ratio on 67P and on Earth explains why the general pattern between gravitational slopes and surface morphologies looks ``familiar'' to our eyes, from high-slope terrains (cliffs) to intermediate-slope terrains (mass wasting as boulder fields) and low-slope terrains (fine material). To first order, gravity shapes terrains in a similar way on 67P and on Earth.

\subsection{Low strengths and surface morphologies related to activity}

The low strength of the cometary material may help to understand some of the surface morphologies, in particular those implying the rising up of a gas bubble, such as the roundish features of 67P \citep[Imhotep region;][]{Auger2015} and 19P \citep{Brownlee2004}, or the depressions adjacent to smooth terrains on the nucleus of 9P \citep{Belton2009}. The saturated vapour pressure of the gas inside the nucleus exceeds 1\,kPa when the temperature exceeds 55\,K for CO and 155\,K for CO$_2$ \citep{Prialnik2004}. This
means that the CO and CO$_2$ gas pressure may locally exceed the compressive strength of the cometary material and the gas may push the surrounding materials while it rises to the surface, creating channels inside the nucleus. This process remains speculative, however, since it has a severe limitation: how would solar energy penetrate meters below the surface, despite the low thermal inertia of 10\,--\,50\,J\,K$^{-1}$\,m$^{-2}$\,s$^{-0.5}$ \citep{Gulkis2015}? We currently do not have the answer to this critical question, but the presence of fractures on the surface \citep{Thomas2015,ElMaarry2015b} may help to solve this issue.

\subsection{Philae observations}

A key question is how to reconcile our low strength estimates with the observation made by the Philae MUPUS experiment, which suggests much larger strengths of several MPa since MUPUS was not able to penetrate the surface material below a few centimeters \citep{Spohn2014}. The solution resides in the hard layer of water ice under the surface dust deposit. As explained in \citet{Pommerol2015}, laboratory experiments with analogs such as KOSI \citep{Grun1991} have shown that a hard layer of water ice can be produced by sublimation/redeposition cycles and/or sintering of water ice close to the surface. This layer has an estimated thickness of a few centimeters \citep{Pommerol2015} to several meters \citep{Kossacki2015} and has been heavily processed, so that it is now hard with strengths up to 1\,MPa as measured by \citet{Jessberger1989}, \citet{Seiferlin1995}, or \citet{Kochan1989}. This layer, which is \textup{\textup{\textup{\emph{not}}}} representative of the bulk nucleus material, provides a simple explanation for the MUPUS observations. The presence of this layer is compatible with our results on strength since we derived i) the tensile strength from large overhangs and collapsed structures, for which this layer is likely negligible, otherwise the measured tensile strength would be much larger, ii) the shear strength from boulders on the surface, for which this buried layer should not matter, and iii) the compressive strength from the tensile strength using mechanical considerations, for which the same justification can be used. 

\subsection{What is representative of the bulk nucleus material?}
 
We derived strengths for different materials in different locations on the nucleus, and the question naturally arises whether these materials are representative of the bulk nucleus material. While it is not possible to give a firm positive answer to this question, we provide here some arguments supporting this idea. 

First, it is very satisfying to see in Table\,1 that strength estimates derived for different materials at different scales and with different methods are mostly consistent with each other, and this from unconsolidated material with modeling at the micrometer scale to consolidated material with observations at the meter scale. When translated to the same scale, dust aggregates, pebbles, boulders, or larger structures have similar strengths. Since they are the building blocks of the nucleus, their mechanical properties should be representative of that of the bulk cometary material. 

Second, we determined the tensile strength for some large collapsed structures, with a thickness of up to 30\,m. These large structures should contain a significant fraction of material that has not been, or has only partially been, affected by activity processes related to insolation. As a result of the low thermal inertia of 10\,--\,50\,J\,K$^{-1}$\,m$^{-2}$\,s$^{-0.5}$ \citep{Gulkis2015}, the thermal heat wave only penetrates the nucleus by a few meters (at most) \citep[e.g.,][]{Groussin2013}. This is confirmed by modeling, which shows that only the first 5 meters are indeed affected by insolation \citep{Prialnik2004}. Complex models also show that water ice is buried under a thin layer (cm) of material depleted in volatiles \citep{Skorov2012}. The presence of water ice close to the surface ($\sim$1\,m) was also observed by the Deep Impact experiment \citep{Sunshine2007}. All these arguments support the idea that activity processes only affect the first meters of the nucleus. The large 30\,m thick structures for which we derived the tensile strength should therefore be a good proxy for the bulk material of the nucleus, even if they are not fully representative of it. We furthermore add that it is currently unclear whether the alteration of the first meters would make the material weaker through forming fractures, or stronger through
forming a hard sintered layer. 

Third, it is important to add that, as demonstrated by \citet{Toth2006}, \citet{Toth2010}, and \citep{Davidsson2001}, even a low tensile strength of a few hundred Pa is amply sufficient to prevent the rotational breakup of the nucleus of 67P with its current rotation period of 12.4\,h. 

Finally, although we currently have no evidence for it, we cannot exclude heterogeneities in strengths over the nucleus. For example, the Hathor region, for which we determined a shear strength $>$1770\,Pa at the 1\,m scale, may be stronger than some small overhangs. Addressing this problem requires more extensive investigations
and is beyond the scope of this paper. For now, we can only argue that the general pattern of Fig.\,\ref{Fig_sketch_slope_terrain}, which is directly linked to the material strengths, is valid across the entire nucleus.

\subsection{Constraining the origin of 67P}

\begin{figure*}
\centering
\includegraphics[width=\hsize]{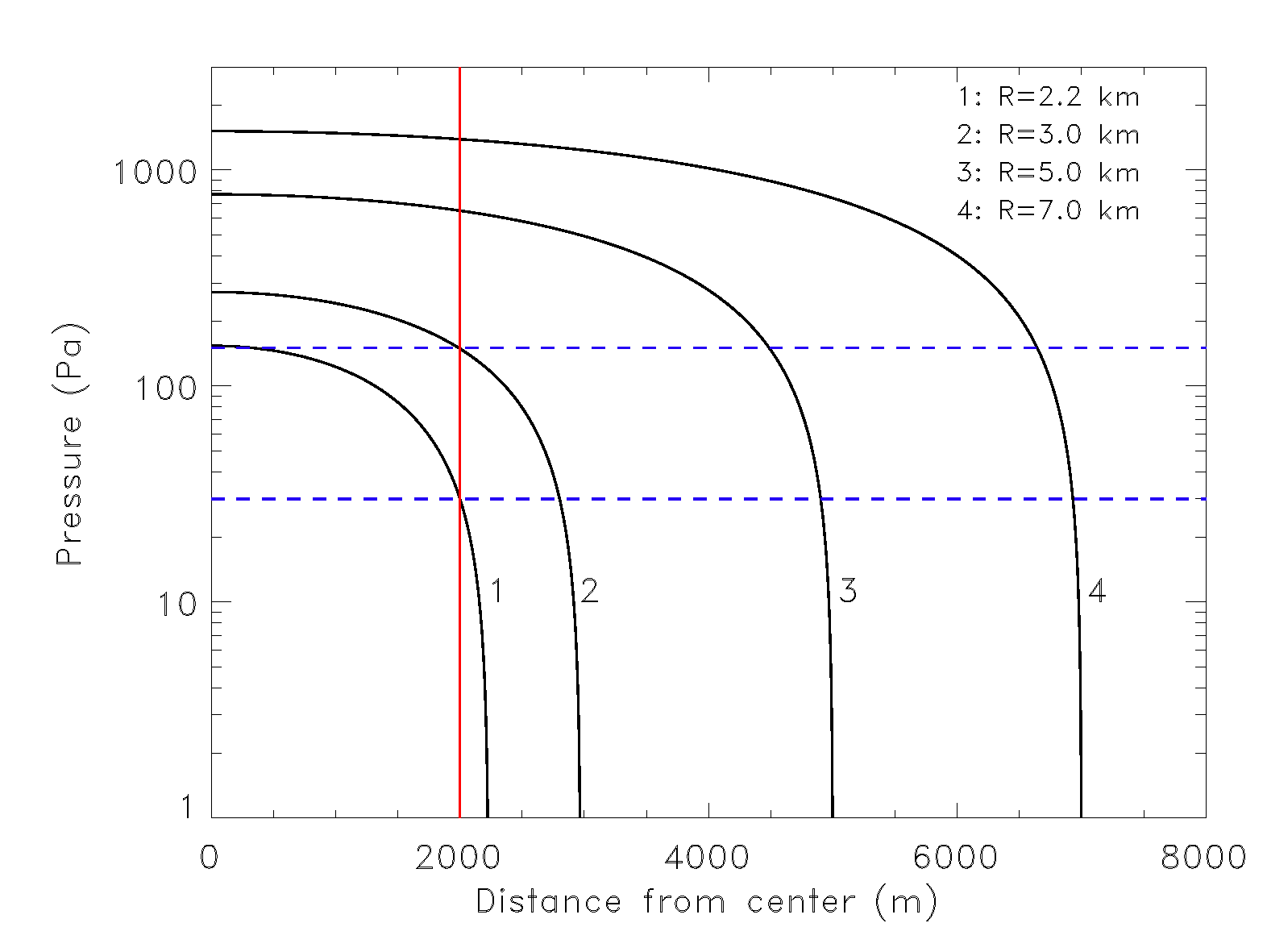}
\caption{Pressure versus depth inside the nucleus, for different initial sizes of the nucleus (radius $R$=2.2, 3.0, 5.0 or 7.0\,km). The pressure increases with depth. The blue range corresponds to the best value of 30-150\,Pa for the compressive strength of the surface materials we see today (30\,--\,150\,Pa). The red line at 2\,km indicates the current size of the nucleus, which is what we see today. For an initial radius of 2.2\,--\,3.0\,km, slightly larger than today, the pressure inside the nucleus exceeds the compressive strength of the layers we see today.}
\label{Fig_pressure_depth}
\end{figure*}

It is interesting to compare the compressive strength to the pressure resulting from gravity inside the nucleus. Assuming a constant density, the pressure $P$ inside the nucleus is given by Eq.\,(\ref{Eq_8}), where $R$ is the nucleus radius and $r$ is the distance from the center.
\begin{equation}
P(r) = \frac{2}{3} \pi G \rho^2 R^2 \bigg[ 1-(\frac{r}{R})^2 \bigg]
\label{Eq_8}
\end{equation}
Figure\,\ref{Fig_pressure_depth} shows the pressure inside the nucleus as a function of depth for different initial radii of the nucleus after its complete accretion and before it entered the inner solar system. Depending on past evolution scenarios, the nucleus of 67P was initially larger by a few hundred meters to a few kilometers \citep{Groussin2007b}. The nucleus has shrunk
as a result of erosion and now has a radius of $\sim$2\,km (red line in Fig.\,\ref{Fig_pressure_depth}). For an initial radius of 2.2\,--\,3.0\,km, slightly larger than today, the pressure inside the nucleus exceeded the compressive strength of the layers we see today (30\,--\,150\,Pa). This means that if the nucleus was formed by the gentle accretion of pristine materials (see below), their subsequent compaction by gravity inside the nucleus was sufficient to explain the compressive strength we measure today.

Continuing with this idea, one may ask whether compaction by gravity inside the nucleus was also sufficient to form the layers we see today on the nucleus \citep{Massironi2015}. On Earth, the diagenesis of rocks starts when they are stressed to more than ten times their compressive strength. If the same mechanism applies to 67P, the primordial material that accreted to form the nucleus had a compressive strength at least ten times lower than the one we measure today to initiate diagenesis, that is, as low as 3\,-- 15\,Pa. In these conditions, compression could have contributed to the formation of layers.

Following the model of \citet{Skorov2012}, the tensile strength of dust aggregates formed by gravitational instability typically
is 1\,Pa. This value is remarkably consistent with our results (Table\,1), particularly if we take into account a compaction scenario where the primordial material was even weaker. Our results therefore tend to favor a formation of comets by pebble accretion in a region of higher concentration of particles such as vortices \citep{Johansen2014,Barge1995,Blum2014}, which implies a gentle formation process by accretion at low velocity, on the order of 1\,m\,s$^{-1}$ or lower. In contrast, the hierarchical accretion model \citep{Weidenschilling2004} with velocities up to 50\,m\,s$^{-1}$ for particles larger than 1\,m, and the collisional scenario between two large bodies of tens of km or more \citep{Davis1997} with an internal compression by gravity larger than 10\,kPa, although not excluded, are less favored. 

Finally, the low tensile strength indicates that the nucleus of 67P very likely did not experience a global melting/freezing of its water content, neither during the accretion stage by radiogenic heating \citep{Podolak2000}, nor more recently by the exothermic amorphous-to-crystalline water ice reaction \citep{Prialnik2004}. Indeed, if a global melting/freezing had occurred, the nucleus would be much harder, from 10\,kPa \citep{Miles2012} to 1\,MPa \citep[e.g.,][]{Seiferlin1995}, and denser with a bulk density higher than 1000\,kg\,m$^{-3}$ \citep{Miles2012}.

\begin{acknowledgements}
OSIRIS was built by a consortium of the Max-Planck-Institut für Sonnensystemforschung, Katlenburg-Lindau, Germany; CISAS University of Padova, Italy; the Laboratoire d'Astrophysique de Marseille, France; the Instituto de Astrofísica de Andalucia, CSIC, Granada, Spain; the Research and Scientific Support Department of the ESA, Noordwijk, Netherlands; the Instituto Nacional de Técnica Aeroespacial, Madrid, Spain; the Universidad Politéchnica de Madrid, Spain; the Department of Physics and Astronomy of Uppsala University, Sweden; and the Institut für Datentechnik und Kommunikationsnetze der Technischen Universität Braunschweig, Germany. The support of the national funding agencies of Germany (DLR), France (CNES), Italy (ASI), Spain (MEC), Sweden (SNSB), and the ESA Technical Directorate is gratefully acknowledged. We thank the Rosetta Science Operations Centre and the Rosetta Mission Operations Centre for the successful rendezvous with comet 67P/Churyumov-Gerasimenko.
We thank David Romeuf from the University Claude Bernard Lyon 1 (France) for creating the red/blue anaglyph in Fig.\,\ref{Fig_Anaglyph}.
We thank the referee, J.\,Blum, for his helpful and constructive report.
\end{acknowledgements}


\bibliographystyle{aa}
\bibliography{References_Strengths}


\begin{table*}
\caption{Summary of the tensile strength ($\sigma_T$), shear strength ($\sigma_S$), and compressive strength ($\sigma_C$) for cometary materials, including our own estimates. The list is not exhaustive, but representative of values found in the literature. We scaled each value to the meter scale using a power law with an exponent equal to -0.6 (Sect.\,1). This scaling should be taken with caution since consolidated and unconsolidated materials may not follow the same scaling law.}
\label{Table_strength}
\begin{tabular}{llrrr}
\hline 
\noalign{\smallskip}
Reference   & Method & Value (Pa) & Scale            & Value (Pa)\\
   &        &       &                  & at 1\,m scale\\
\noalign{\smallskip}
\hline 
\noalign{\smallskip}
{\bf Tensile strength ($\sigma_T$)}\\
\noalign{\smallskip}
\citet{Klinger1989}     & Observations (sungrazing comets) & 100                  & 1\,m         & 100 \\
\citet{Davidsson2001}   & Observations (rotational breakup)& 1\,--\,53            & 1\,m         & 1\,--\,53\\
\citet{Asphaug1996}   & Observations (D/1993\,F2 SL9)      & 5                    & 1\,m         & 5 \\
\citet{Blum2006}        & Laboratory experiments           & 200\,--\,1\,100      & 1\,cm        & 13\,--\,69 \\
\cite{BarNun2007}       & Laboratory experiments           & 2\,000\,--\,4\,000   & 100\,$\mu$m  & 8\,--\,16 \\
\citet{Blum2014}        & Laboratory experiments           & 1                    & 1\,mm        & 0.02\\
\citet{Biele2009}       & Modelling                         & 5\,000               & 10\,$\mu$m   & 5 \\
\citet{Greenberg1995}   & Modelling                         & 270                  & 10\,$\mu$m   & 0.3 \\
\noalign{\smallskip}
\noalign{\smallskip}
{\it \underline{This work}}         & Observations (67P, Rosetta)             &             &          &  \\
                                    & \hspace{0.3cm}-- Collapsed structures   & $<$150      & 30\,m           & $<$1\,150 \\
                                    & \hspace{0.3cm}-- Overhangs              & 3\,--\,15   & 5\,m            & 8\,--\,39  \\

\noalign{\smallskip}
\noalign{\smallskip}
\hline 
\noalign{\smallskip}
{\bf Shear strength ($\sigma_S$)}\\
\noalign{\smallskip}
\citet{Holsapple2007}  & Observations (9P, Deep Impact)          & 0\,--\,12\,000  & 1\,m         & 0\,--\,12\,000 \\
\citet{AHearn2005}        & Observations (9P, Deep Impact)       & 65              & 1\,m         & 65 \\
\noalign{\smallskip}
\noalign{\smallskip}
{\it \underline{This work}}          & Observations (67P, Rosetta)                           &             &        &  \\
                                     & \hspace{0.3cm}-- Hathor cliffs                        & $>$30       &900\,m  & $>$1\,770 \\
                                     & \hspace{0.3cm}-- Fine surface materials and boulders  & 4\,--\,30   &1\,m    & 4\,--\,30  \\

\noalign{\smallskip}
\noalign{\smallskip}
\hline 
\noalign{\smallskip}
{\bf Compressive strength ($\sigma_C$)}\\
\noalign{\smallskip}
\citet{Jessberger1989}  & Laboratory experiments            & 30\,000\,--\,1\,000\,000  & 1\,cm        & 1\,890\,--\,63\,100 \\
\citet{BarNun2007}      & Laboratory experiments            & 20\,000                   & 100\,$\mu$m  & 80 \\
\citet{Guttler2009}     & Laboratory experiments            & $<$1\,000                 & 1\,mm        & 16 \\
\citet{Blum2014}        & Laboratory experiments            & 15                        & 1\,mm        & 0.2 \\
\citet{Sirono2000}      & Modelling                          & 6\,500                    & 10\,$\mu$m   & 7 \\
\noalign{\smallskip}
\noalign{\smallskip}
{\it \underline{This work}}     & Observations (67P, Rosetta)           &             &          &  \\
                               & \hspace{0.3cm}-- Fine surface materials  & $<$15\,600   &10\,cm            & $<$3\,920   \\
                               & \hspace{0.3cm}-- Collapsed structures    & $<$1\,500    &30\,m            & $<$11\,600 \\         
                                & \hspace{0.3cm}-- Overhangs              & 30\,--\,150  &5\,m            & 79\,--\,394 \\
                       
 \noalign{\smallskip}
\noalign{\smallskip}
\hline 
\end{tabular}
\end{table*}


\end{document}